\documentclass[longauth]{aa}
\usepackage{graphicx}
\usepackage{txfonts}

\usepackage{pdflscape}
\usepackage{xcolor}
\usepackage{lscape}
\usepackage{float}
\usepackage[hidelinks]{hyperref}
   \hypersetup{
      colorlinks = true,
      citecolor ={cyan}
   }
\usepackage{longtable}
\newcommand{\RVpp} {RV$_{\rm PP}$~}
\def\kms{km\,s$^{-1}$}

\newcommand{\Msun}{$\,M_\odot$\xspace}

\begin{document}

   \title{Binarity at LOw Metallicity (BLOeM)\thanks{Based on observations collected at the European Southern Observatory under ESO program ID 112.25W2.} \thanks{Table \ref{table:all_bsgs} is also available in electronic form at the CDS via anonymous ftp to cdsarc.u-strasbg.fr (130.79.128.5) or via http://cdsweb.u-strasbg.fr/cgi-bin/qcat?J/A+A/}}

   \subtitle{Multiplicity of early B-type supergiants in the Small Magellanic Cloud}

   \author{N.\ Britavskiy \inst{\ref{inst:rob},\ref{inst:liege}}
    \and L.\ Mahy \inst{\ref{inst:rob}}
    \and D.\ J.\ Lennon\inst{\ref{inst:iac}, \ref{inst:ull}} 
    \and L.\ R.\ Patrick\inst{\ref{inst:cab}}
    \and H.\ Sana\inst{\ref{inst:kul}}
    \and J.\ I.\ Villase\~{n}or\inst{\ref{inst:mpia}}
    \and T.\ Shenar\inst{\ref{inst:TelAv}}
    \and J.\ Bodensteiner\inst{\ref{inst:eso}} 
    \and M.\ Bernini-Peron\inst{\ref{inst:ari}}
    \and S.\ R.\ Berlanas\inst{\ref{inst:iac}, \ref{inst:ull}}
    \and D.~M.~Bowman\inst{\ref{inst:newcastle}, \ref{inst:kul}}
    \and P.\ A.\ Crowther\inst{\ref{inst:sheffield}} 
    \and S.\ E.\ de Mink \inst{\ref{inst:mpa}}
    \and C. J. Evans \inst{\ref{inst:esa_stsi}}
    \and Y.\ G{\"o}tberg \inst{\ref{inst:ista}}
    \and G.\ Holgado\inst{\ref{inst:iac}, \ref{inst:ull}}
    \and C.\ Johnston\inst{\ref{inst:mpa}}
    \and Z.~Keszthelyi\inst{\ref{inst:naoj}}
    \and J.\ Klencki\inst{\ref{inst:eso}} 
    \and N.\ Langer\inst{\ref{inst:bonn}}
    \and I.\ Mandel\inst{\ref{inst:monash},\ref{inst:ozgrav}}
    \and A.\ Menon\inst{\ref{inst:columbia}}
    \and M.\ Moe\inst{\ref{inst:wyoming}} 
    \and L.\ M.\ Oskinova\inst{\ref{inst:up}}
    \and
    D. Pauli\inst{\ref{inst:up},\ref{inst:kul}}
    \and M.\ Pawlak\inst{\ref{inst:lund}}
    \and V.\ Ramachandran\inst{\ref{inst:ari}}
    \and M.\ Renzo\inst{\ref{inst:AZ}}
    \and A.\ A.\ C.\ Sander\inst{\ref{inst:ari}}
    \and F.\ R.\ N.\ Schneider\inst{\ref{inst:hits},\ref{inst:ari}}
    \and A.\ Schootemeijer\inst{\ref{inst:bonn}}
    \and K.\ Sen\inst{\ref{inst:umk},\ref{inst:AZ}}
    \and S.\ Sim\'on-D\'iaz\inst{\ref{inst:iac}, \ref{inst:ull}} 
    \and J.\ Th.\ van Loon\inst{\ref{inst:keele}}
    \and J.\ S. Vink\inst{\ref{inst:armagh}}
 }

\institute{
{Royal Observatory of Belgium, Avenue Circulaire/Ringlaan 3, B-1180 Brussels, Belgium \label{inst:rob}}
\and
{University of Li\`ege, All\'ee du 6 Ao\^ut 19c (B5C), B-4000 Sart Tilman, Li\`ege, Belgium \label{inst:liege}}
\and    
{Instituto de Astrof\'{\i}sica de Canarias, C. V\'{\i}a L\'actea, s/n, 38205 La Laguna, Santa Cruz de Tenerife, Spain\label{inst:iac}}
\and    
{Universidad de La Laguna, Dpto. Astrof\'{\i}sica, Av.\ Astrof\'{\i}sico Francisco S\'anchez, 38206 La Laguna, Santa Cruz de Tenerife, Spain\label{inst:ull}}
\and
{Centro de Astrobiolog\'{\i}a (CSIC-INTA), Ctra.\ Torrej\'on a Ajalvir km 4, 28850 Torrej\'on de Ardoz, Spain\label{inst:cab}}
\and
{Institute of Astronomy, KU Leuven, Celestijnenlaan 200D, 3001 Leuven, Belgium\label{inst:kul}}
\and
{Max-Planck-Institut f\"{u}r Astronomie, K\"{o}nigstuhl 17, D-69117 Heidelberg, Germany\label{inst:mpia}}
\and
{The School of Physics and Astronomy, Tel Aviv University, Tel Aviv 6997801, Israel\label{inst:TelAv}}  
   \and
{ESO - European Southern Observatory, Karl-Schwarzschild-Strasse 2, 85748 Garching bei M\"unchen,
Germany \label{inst:eso}}
\and
{Zentrum f\"ur Astronomie der Universit\"at Heidelberg, Astronomisches Rechen-Institut, M\"onchhofstr. 12-14, 69120 Heidelberg, Germany\label{inst:ari}} 
\and 
{School of Mathematics, Statistics and Physics, Newcastle University, Newcastle upon Tyne, NE1 7RU, UK\label{inst:newcastle}}
\and
{Department of Physics \& Astronomy, Hounsfield Road, University of Sheffield, Sheffield, S3 7RH, United Kingdom\label{inst:sheffield}}
\and {Max-Planck-Institute for Astrophysics, Karl-Schwarzschild-Strasse 1, 85748 Garching, Germany\label{inst:mpa}}
\and {European Space Agency (ESA), ESA Office, Space Telescope Science Institute, 3700 San Martin Drive, Baltimore, MD 21218, USA}\label{inst:esa_stsi}
\and
{Institute of Science and Technology Austria (ISTA), Am Campus 1, 3400 Klosterneuburg, Austria\label{inst:ista}}
\and
{Center for Computational Astrophysics, Division of Science, National Astronomical Observatory of Japan, 2-21-1, Osawa, Mitaka, Tokyo 181-8588, Japan\label{inst:naoj}}
\and
{Argelander-Institut f\"{u}r Astronomie, Universit\"{a}t Bonn, Auf dem H\"{u}gel 71, 53121 Bonn, Germany\label{inst:bonn}}
\and
{School of Physics and Astronomy, Monash University, Clayton VIC 3800, Australia\label{inst:monash}}
\and
{ARC Centre of Excellence for Gravitational-wave Discovery (OzGrav), Melbourne, Australia\label{inst:ozgrav}}
\and
{{Department of Astronomy, 538 West 120th Street, Pupin Hall, Columbia University, New York City, NY 10027, U.S.A}\label{inst:columbia}}
\and
{Department of Physics and Astronomy, University of Wyoming, 1000 E. University Ave., Dept. 3905, Laramie, WY 82071, USA\label{inst:wyoming}}
\and
{Institut f\"ur Physik und Astronomie, Universit\"at Potsdam, Karl-Liebknecht-Str. 24/25, 14476 Potsdam, Germany\label{inst:up}}
\and {Lund Observatory, Division of Astrophysics, Department of Physics, Lund University, Box 43, SE-221 00, Lund, Sweden}\label{inst:lund}
\and
{Department of Astronomy \& Steward Observatory, 933 N. Cherry Ave., Tucson, AZ 85721, USA\label{inst:AZ}}
\and
{Heidelberger Institut f{\"u}r Theoretische Studien, Schloss-Wolfsbrunnenweg 35, 69118 Heidelberg, Germany\label{inst:hits}}
\and {Institute of Astronomy, Faculty of Physics, Astronomy and Informatics, Nicolaus Copernicus University, Grudziadzka 5, 87-100 Torun, Poland\label{inst:umk}}
\and
{Lennard-Jones Laboratories, Keele University, Newcastle, ST5 5BG, UK\label{inst:keele}}
\and 
{Armagh Observatory, College Hill, Armagh, BT61 9DG, Northern Ireland, UK\label{inst:armagh}}
}
\offprints{mbritavskiy@uliege.be}

\date{Submitted 11 November 2024 / Accepted 4 February 2025}

 
  \abstract
   {The blue supergiant (BSG) domain contains a large variety of stars whose past and future evolutionary paths are still highly uncertain.  Since binary interaction plays a crucial role in the fate of massive stars, investigating the multiplicity among BSGs helps shed light on the fate of such objects.}
   {We aim to estimate the binary fraction of a large sample of BSGs in the Small Magellanic Cloud (SMC) within the Binarity at LOw Metallicity (BLOeM) survey. In total, we selected 262 targets with spectral types B0-B3 and luminosity classes I-II.}
   {This work is based on spectroscopic data collected by the {\sc giraffe} instrument, mounted on the Very Large Telescope, which gathered nine epochs over three months. Our spectroscopic analysis for each target includes the individual and peak-to-peak radial velocity measurements, an investigation of the line profile variability, and a periodogram analysis to search for possible short- and long-period binaries. }
   {By applying a 20 \kms~ threshold on the peak-to-peak radial velocities above which we would consider the star to be binary, the resulting observed spectroscopic binary fraction for our BSG sample is 23 $\pm$ 3$\%$. An independent analysis of line profile variability reveals 11 (plus 5 candidates) double-lined spectroscopic binaries and 32 (plus 41 candidates) single-lined spectroscopic binaries.  
   Based on these results, we estimated the overall observed binary fraction in this sample to be 34 $\pm$ 3$\%$, which is close to the computed intrinsic binary fraction of 40 $\pm$ 4$\%$. In addition, we derived reliable orbital periods for 41 spectroscopic binaries and potential binary candidates, among which there are 17 eclipsing binaries, including 20 SB1 and SB2 systems with periods of less than 10 days. We reported a significant drop in the binary fraction of BSGs with spectral types later than B2 and effective temperatures less than 18 kK, which could indicate the end of the main sequence phase in this temperature regime. We found no metallicity dependence in the binary fraction of BSGs, compared to existing spectroscopic surveys of the Galaxy and Large Magellanic Cloud.} 
   {}


   \titlerunning{Multiplicity of BSGs in the SMC}
   \authorrunning{N. Britavskiy et al.}

   \maketitle
   \nolinenumbers
%

\section{Introduction}
B-type supergiants (BSGs),  also simply referred to as "blue supergiants," comprise a major piece of the puzzle  in our understanding of massive star evolution. From a spectral morphology point of view, BSGs exhibit the luminosity class of (LC) Iab/II, which corresponds to the bolometric magnitude of $<$ -7 mag, and they have specific diagnostic metallic (mainly silicon and magnesium) absorption lines in the spectra \cite[][]{Lennon_1997,Negueruela_2024}.
Given their basic parameters and position in the Hertzsprung–Russell diagram (HRD), they are located in a region that is predicted to mark the transition between core hydrogen and helium-burning. The precise location of this region depends a great deal on our understanding of the processes in the stellar interior, which are still subject to severe uncertainties, such as internal mixing \citep[for a review of how mixing influences the fate of BSGs, see][]{Bowman_2020}. The same HRD region can be covered by objects with different internal structures, such as more evolved helium-burning stars that have lost a part of their outer layer, either intrinsically or via binary interaction. Moreover, there are serious open questions concerning the surface and wind properties of B-type supergiants and their evolutionary origins \citep[see][]{Trundle_2004, Bernini-Peron_2024}.



On the main sequence (MS), the progenitors of  BSGs are OB-type stars. By now, it has been well established that OB stars have high multiplicity rates \citep[50-70$\%$, see][]{Sana_2012,Banyard_2022,Offner_2022}. 
However, for many decades, it was widely  believed that BSGs are preferably single objects \citep{Humphreys_1978,McEvoy_2015,bsgs_iacob,de_Burgos_2024}. Thus, we would naturally expect that a significant fraction of BSGs are products of binary interaction. One of the explanations could be that, in the BSG regime, there is a high fraction of merger products \citep{Vanbeveren_2013,Justham_2014,Menon_2024,Schneider_2024}. One of the most famous example is the progenitor star of SN 1987A, which was expected to be a red supergiant \citep[RSG, see][]{Walborn_1987ApJ}; however, it was later shown  the progenitor is a BSG and a binary-merger scenario provides a natural explanation for both the BSG star and the rings of ejected material surrounding it \citep{Podsiadlowski_1992PASP,Podsiadlowski_2007A,Menon_2017}. 
Another hint of predominance of binary interaction products in the BSG domain is a low multiplicity fraction of BSGs claimed in different multi-epoch spectroscopic surveys across different metallicity environments, namely, the estimated observed binary fraction of BSGs in the Galaxy is  $\sim$27$\%$ \citep[see,][]{burgos_2025} and in the Large Magellanic Cloud (LMC), it is approximately 23 $\pm$ 6$\%$ \citep{Dunstall_2015}. 

The existence of blue loop phenomena challenges our understanding of BSG even further because it depends sensitively on the adopted core and envelope boundary physics \citep[e.g.,][]{Stothers_1992,Walmswell_2015, Bowman_2019, Farrell_2022}, which remains uncertain. For example, by increasing the semiconvection efficiency, it is possible to keep the star bluer until later in its evolution \citep[as shown in the example of SN 1987A,][]{Podsiadlowski_1989,Langer_1989}. However, this parameter may strongly depend on the metallicity and age of the star. While bona fide blue loops following a RSG excursion are not expected above $\sim$15\,$M_\odot$ \citep[][]{Saio_2013, Walmswell_2015, Schootemeijer_2019, Klencki_2020,Schneider_2024}, secondary stars in interacting binaries where accretion leads to the growth of the convective core during the MS phase \citep[e.g.,][]{Neo_1977,Hellings_1983, Renzo_2023, Wagg_2024} may experience long-lived blue loops \citep[possibly after being ejected from the binary at the core-collapse of their companion; see][]{Renzo_2023}, which could also explain such a low observed spectroscopic binary fraction in this domain.

Thus, BSGs are the ideal targets to study the late evolutionary stages of massive binary systems.
As one of the parameters we can use to investigate them, we can choose their CNO abundances \citep[e.g.,][]{Evans_2004,McEvoy_2015,Mahy_2016, de_Burgos_2024} or we can characterize and reproduce the orbital parameters of known BSG binaries by using evolutionary models if the initial conditions and proper physics of binary interaction are accounted for.  
The existence of short- and long-period BSG binaries may suggest different evolutionary paths that could form such systems \citep[first identified by][]{Stothers_1970}. 
For example, the short-period systems with a BSG-like spectrum can allow us to infer that they are donor stars in the Algol system \citep[case A mass transfer, see e.g.][]{Sen_2022}; whereas BSGs on a long orbit can either be evolved single star (pre-interaction system) or merger products in a triple system.

In addition, it is important to distinguish the origin of LC I BSGs from that of LC II. The less luminous BSGs (LC II) could represent the end of the MS phase of B-type stars, while more luminous BSGs could be post-interaction long-period systems or merger products. Moreover, BSGs of luminosity class Iab could have previously experienced the RSG phase; if the binary system survives a significant radial expansion during this phase, the resulting system should also be a long-period one. 
Indeed, if at some point, BSGs would observationally appear as luminous blue variables, 60$\%$ of them would be expected to be in long-period binary systems  \citep{Mahy_2022lbv}.
Thus, once we are able to constrain the different evolutionary paths of BSG origins, it will be easier to characterize the overall BSG appearance on the observational HRD; for instance, it would allow us to solve the known problem of the Hertzsprung gap \citep[i.e.\ why we observe a significant number of BSGs relative to what theoretical models predict; see][]{Fitzpatrick_1990,Wesmayer_2022}.

To address these problems, we study the multiplicity properties of massive BSGs ($M_{\rm ini}\gtrsim8$ \Msun) in the Small Magellanic Cloud (SMC) within the new Binarity at LOw Metallicity \citep[BLOeM,][]{bloem_i} survey. Apart from homogeneous and statistically significant numbers of observed massive stars, this survey gives the unique opportunity to study the evolution of BSGs at low metallicity ($Z = 0.2\,Z_\odot$). The BLOeM survey contains multi-epoch spectroscopic observations of nearly 1000 stars in the SMC across different spectral types. The survey has been divided into different sub-samples that represent various evolutionary stages of massive stars across the HRD, namely: the O-type star domain (Sana et al., subm. in Nature Astronomy), the B-type dwarfs \citep[spectral types B0-B3, see][]{Villasenor_2025}, OeBe stars \citep{Bodensteiner_2025}, the BSGs (present work), and the cool supergiants \citep[spectral types later than B3 up to F-type supergiants, see][]{Patrick_2025}. 

The main goal of the present paper is to provide the first preliminary multiplicity constraints of the large and uniform sample of BSGs located in the SMC. The paper has the following structure: In Sect. 2, we present briefly the available dataset. In Sect. 3, we describe in detail all steps of radial velocity (RV) determination and spectroscopic binary classification. Sections 4 and 5 present a discussion and conclusions, respectively.
\section{Sample selection and observations}

The entire BLOeM sample was built by using the third Gaia data release catalog \citep{Gaia2023DR3}, combined with a parallax and proper motion filters to eliminate the contamination from foreground objects. Then, to select only massive stars, we used the SMC evolutionary tracks \citep{Schootemeijer_2019} and appropriate colour cuts. More detailed information about the sample selection, spectral type (SpT) classification, and data reduction can be found in \citet{bloem_i}. The BLOeM ESO observing proposal is assembling  25 epochs of observations over a two-year timespan; however, the present work is based on the nine first epochs of observations spread over three months in 2023.  
The spectra were obtained with the VLT FLAMES/GIRAFFE spectrograph and cover the wavelength range 3950 to 4560 \AA\  with a spectral resolving power of $R=\lambda/\Delta \lambda=6200$. The individual spectra have a signal-to-noise ratio per pixel (S/N) that ranges from $\sim$30 up to 300, with a median value ranging from 70 to 100 depending on the observed field.

\begin{figure*}[!ht]
\centering
\includegraphics[width=0.75\textwidth]{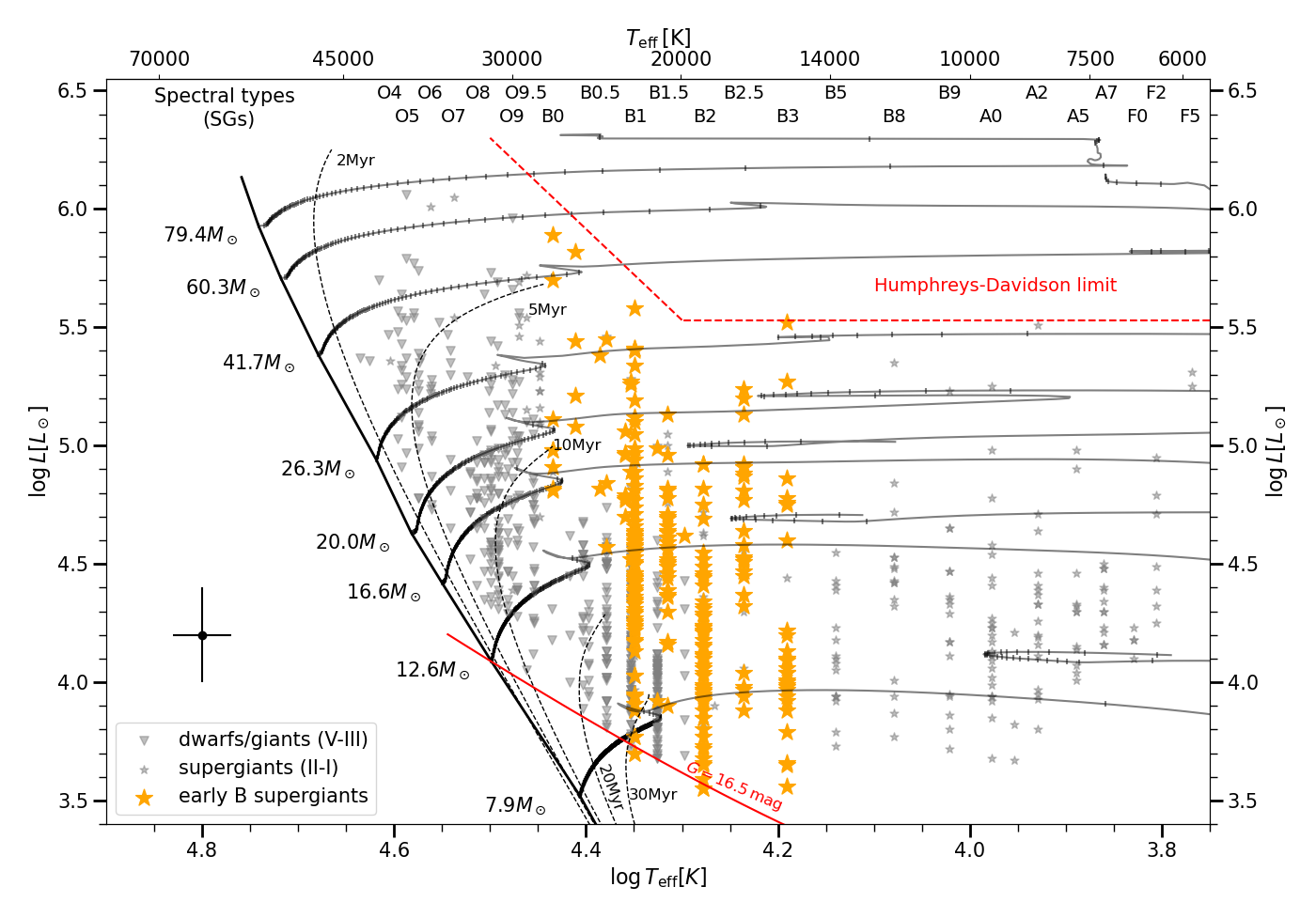}
\caption{Hertzprung-Russell diagram for the entire BLOeM survey with highlighted early B-type supergiants that have been selected for study in the present work (marked with orange star symbols, 262 in total). Non-rotating stellar evolutionary tracks represent the evolution of single stars at SMC metallicity calculated by \citet{Schootemeijer_2019} with mass-dependent overshooting \citep[described in Appendix B of][]{Hastings_2021}.}
\label{main_hrd}
\end{figure*}

\begin{figure}[!ht]
\centering
\includegraphics[width=0.49\textwidth]{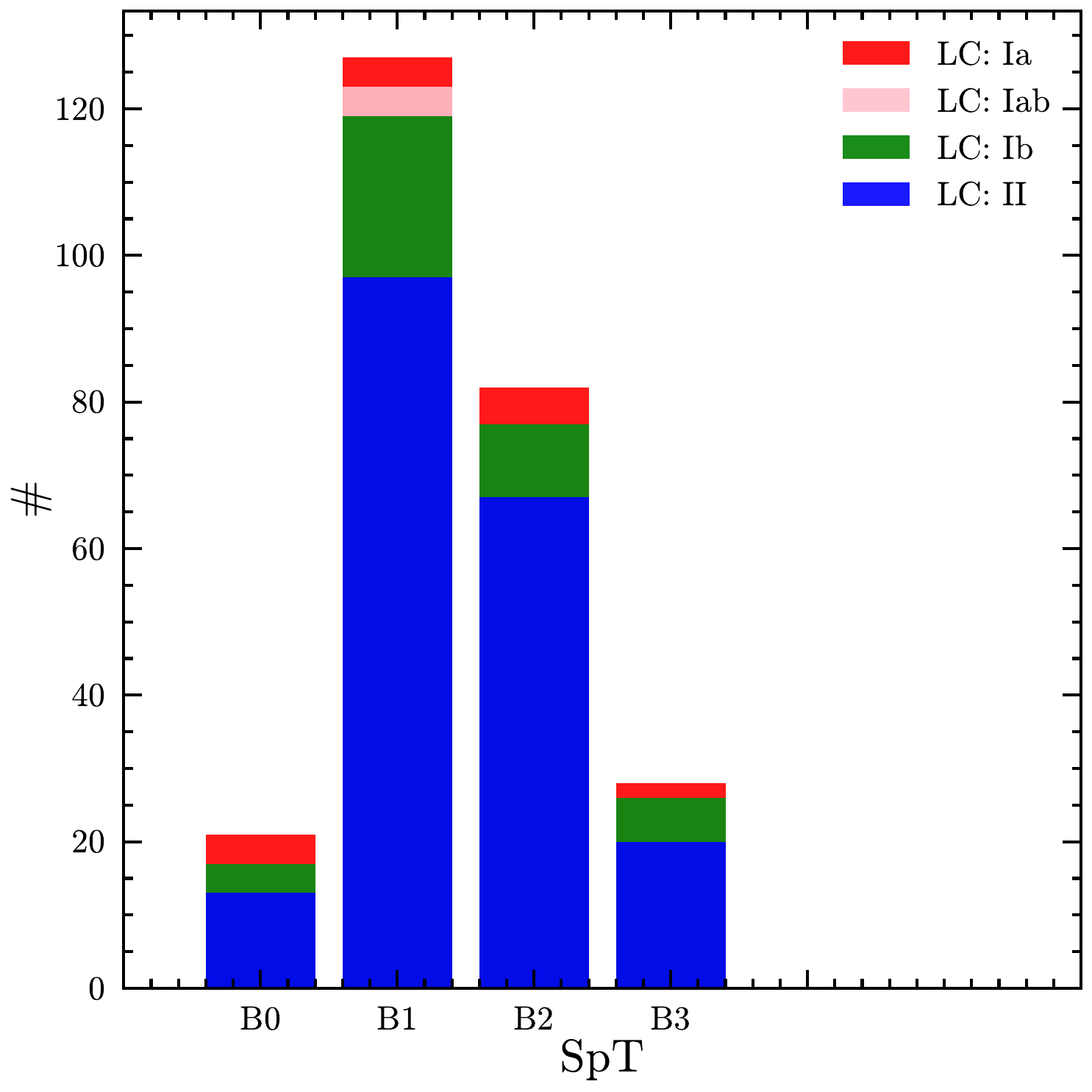}
\caption{Number of selected early B-type supergiants in the BLOeM sample as a function of spectral sub-types and luminosity classes.}
\label{sample_spt_distribution}
\end{figure}

The sample of stars studied in this work is defined as targets with spectral types B0-B3 and luminosity classes I-II, which amounts to 264 targets. However, for two of them (BLOeM 7-066 and BLOeM 7-077), only one epoch was acquired due to instrumental issues;  therefore, we excluded them from the analysis so far. The position of the BLOeM targets on the HRD is shown in Fig.~\ref{main_hrd}. In addition, in Fig.~\ref{sample_spt_distribution}, we present the distribution of the final sample of BSGs (262 targets in total) by spectral types and luminosity class. With the preliminary estimates of effective temperatures ($T_{\rm eff}$) and luminosities (log($L/L_\odot$)) taken from \citet{bloem_i} we also estimated the values of radii (R) using the Stefan-Boltzmann law (we listed all fundamental parameters for each BSG in Table \ref{table:all_bsgs}). The values of $T_{\rm eff}$ were determined by using the SMC SpT-temperature relation \citep[][]{Dufton_2019} and bolometric luminosities were calculated through the $K_{s}$ band photometry \citep[taken from the VISTA Magellanic Survey,][]{Cioni_2011}, V-$K_{s}$ colors, and V-band bolometric corrections \citep[][]{Lanz_2007} with the adopted interstellar extinction equal to zero. The extinction in infrared bands can be assumed negligible because the extinction in the V-band towards SMC A$_{V}$ $\sim$0.35  \citep[according to][]{Schootemeijer_2021} that corresponds to A$_{Ks}$ $\sim$0.03. The step in our SpT classification is 0.2-0.3 in the B0 sub-type and 0.5 in later spectral sub-types (i.e.,\ B1, B1.5, etc.), which explains the steps and error bars of $T_{\rm eff}$ values on the HRD (see Fig.~\ref{main_hrd}).

Notably, the majority of BSGs in our sample exhibit LC II, however, the entire sample has a broad luminosity range ($\sim$3.5 < $\log$ (L/L$_{\odot}$) < ~$\sim$5.5) that could overlap with the B-type giants LC: III. The reason for it is in the limited wavelength spectral range presently available, for a more precise LC classification, it is necessary to have a wider wavelength coverage for all BLOeM targets (including the H$\alpha$ region). Thus, for any further interpretation of the results, we prefer to use the different domains of luminosity ranges rather than LCs. At the present stage, the luminosity estimates are more reliable (especially taking into account negligible interstellar extinction towards the SMC). In any case, the given sample of BSGs represents an early B-type supergiant phase that is expected to be evolutionary descendants of the OB-type dwarfs and giants on the MS. We defined the spectral type boundary of the sample as B0 -- B3 because on later spectral types RV variations of supergiants are much smaller \citep[$<$10 \kms, see][]{Patrick_2025}. Thus, the techniques of studying multiplicity properties of late-B, A-, and F-type supergiants should be different. Thus, to have a uniform sample of BSGs in terms of spectral morphology and RV variations we limited our sample to the B3 spectral type.

\section{Analysis methodology}

To get meaningful and comprehensive results regarding the multiplicity of our sample, we applied different approaches that complement each other. First, we estimated the number of BSGs with the amplitude of RV variations above the threshold of 20 \kms. In this way, we derived the observed fraction of spectroscopic binaries in our sample of BSGs. Then we evaluated the line profile variability of all targets to detect potential spectroscopic binaries and BSGs with a significant intrinsic variability. The next steps were to derive reliable orbital periods and evaluate the Roche-lobe filling factor for all targets in our sample. In addition, we estimated the intrinsic binary fraction for BSGs in our sample by correcting the observed spectroscopic binary fraction for the observational biases. In the next subsections, we describe the individual steps of our analysis in detail .

\subsection{Radial velocity measurements}

In principle, BSGs are located in a temperature regime where a considerable number of metal lines (e.g., ions of carbon, silicon, oxygen, and nitrogen) can be found in the optical spectrum, which would mark them as ideal features for RV measurements. However, taking into account the resolution and resulting S/N of the spectroscopic data, these metallic lines are not always prominent. In addition, at the SMC metallicity, all metallic lines are expected to be not as prominent as at the solar metallicity. Thus, to get the individual RV measurements, we took the whole available wavelength range ($\lambda\lambda$3985 -- 4560 \AA) in each spectrum and cross-correlated it with a template. We followed the cross-correlation technique described in \citet{Zucker_2003}. 
As a template, we used the stacked spectra of all available epochs per individual object, which were shifted to the vacuum-wavelength by using TLUSTY model templates. 

As we are performing cross-correlation of the observed spectra with the stacked one for a given star the resulting cross-correlation function (CCF) could have artifacts that are related with a noise auto-correlation or the presence of emission lines. It could affect the morphology of the peak of CCF from which we get the final RV estimates (we present the example of such case in Fig.~\ref{ccf_plot}). To minimize such effects we fit the CCF with a second-degree polynomial function to smooth the final CCF curve. This fitting makes the resulting RV measurements more independent from the S/N, however, it increases the standard deviation of RV estimates because, in the \citeauthor{Zucker_2003} formalism, the final error bars are dependent on the sharpness of the peak in the CCF and the curvature of the parabola is smoother than the original CCF (see the corresponding uncertainties of RV estimates in Fig.~\ref{ccf_plot}). 

\begin{figure*}[!ht]
\centering
\includegraphics[width=0.49\textwidth]{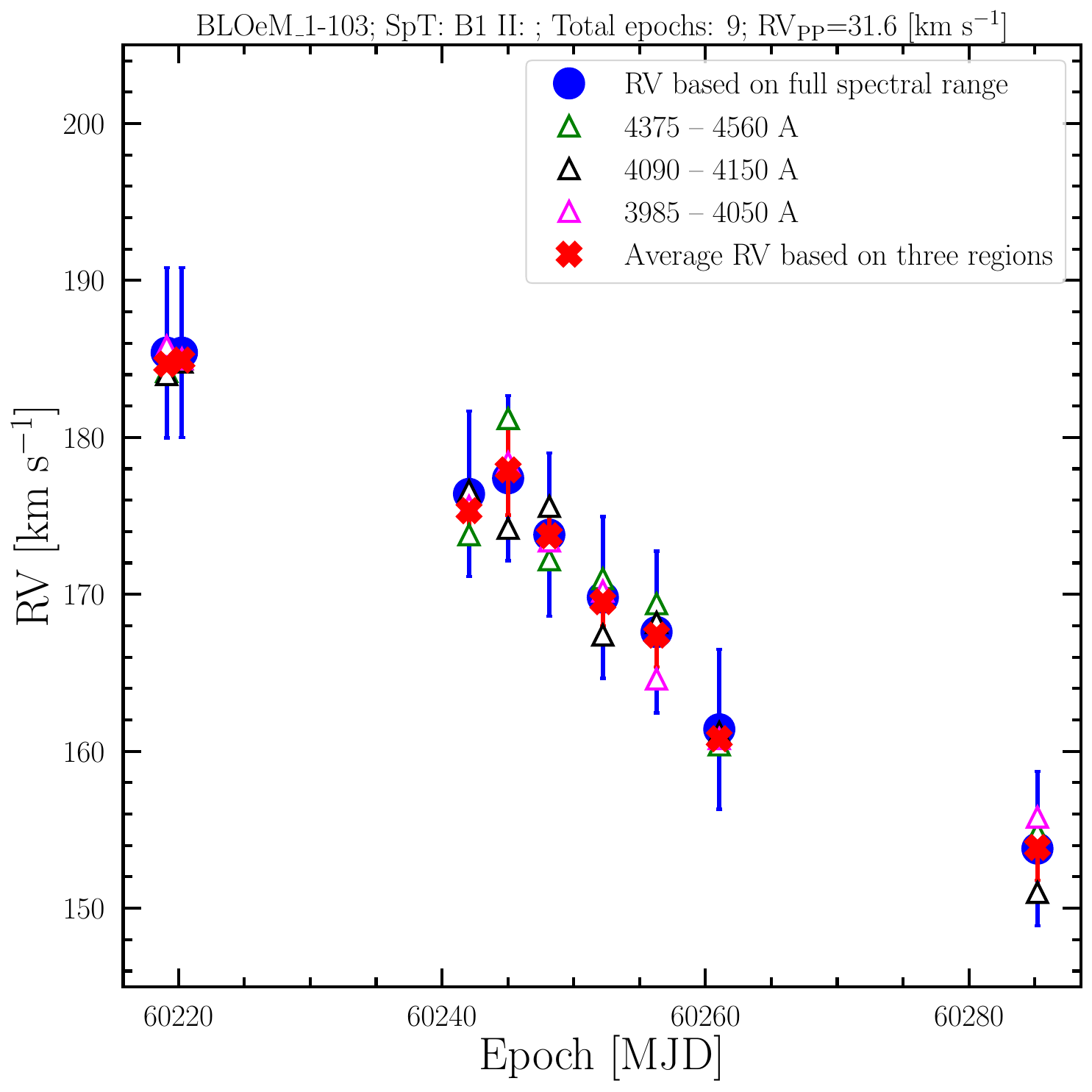}
\includegraphics[width=0.49\textwidth]{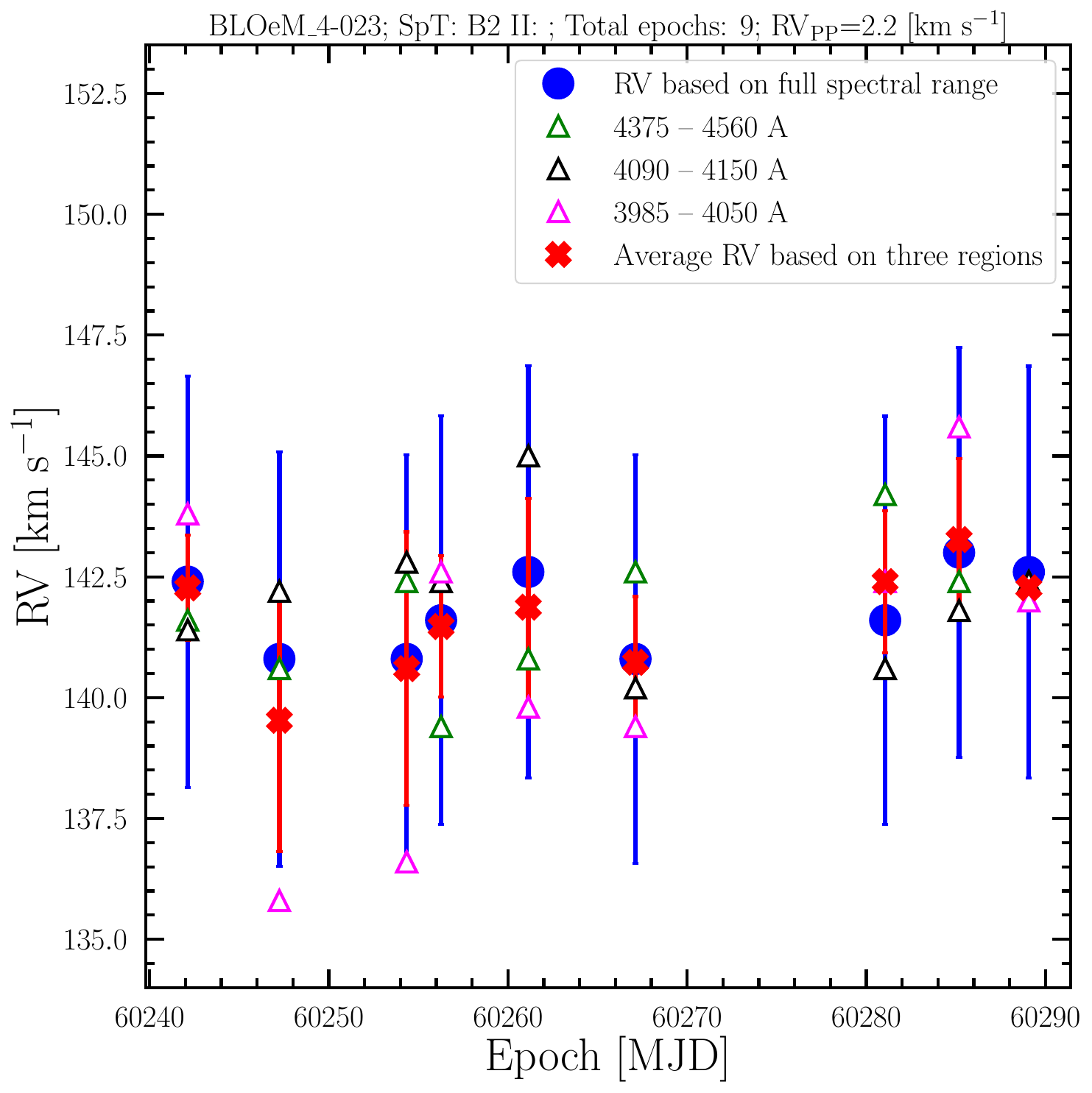}
\caption{Example of individual radial velocity measurements by using a cross-correlation technique based on different wavelength ranges for the SB1 system (left panel, BLOeM 1-103) and an apparently single BSG with small RV variation (right panel: BLOeM 4-023, noting that the ordinate axis is significantly zoomed-in). Blue error bars represent the standard deviation of measurements based on the cross-correlation of the entire wavelength range and the red error bars represent the standard deviation of mean radial velocity based on the three different wavelength pieces.}
\label{rv_examples}
\end{figure*}

In addition, to check how the wavelength calibration remains stable throughout the entire spectrum range and to investigate the contribution of individual metallic lines to the resulting RV measurements, we decided to cut the spectrum into three segments and perform cross-correlation with respect to the stacked reference spectrum accordingly. Specifically, we chose the following segments: ({\it i}) $\lambda\lambda$3985 -- 4050 \AA~ (with present N\,{\sc ii}\,$\lambda 3995$, He\,{\sc i}\,$\lambda 4009$, He\,{\sc i}\,$\lambda 4026$ lines), ({\it ii}) $\lambda\lambda$4090 -- 4150 \AA~  (H\,{\sc $\delta$}\,$\lambda 4100$, Si\,{\sc iv}\,$\lambda 4116$, Si\,{\sc ii}\,$\lambda\lambda 4128-4130$ lines available), and ({\it iii}) $\lambda\lambda$4375 -- 4560 \AA~ (with a prominent He\,{\sc i}\,$\lambda 4387$, He\,{\sc i}\,$\lambda 4471$, Mg\,{\sc ii}\,$\lambda 4481$, and Si\,{\sc iii}\,$\lambda 4552$ lines). The results of these measurements are presented in Fig.~\ref{rv_examples} on an example of objects with significant (BLOeM 1-103) and extremely small (BLOeM 4-023) RV variations.
The main conclusion of this exercise is that RV measurements based on the three different ranges are within the standard deviation of measurements based on the cross-correlation of the entire wavelength range (blue error bars in Fig.~\ref{rv_examples}). 
Moreover, the presence of prominent hydrogen and helium lines even in low S/N spectra makes cross-correlation results of the entire spectrum more reliable with respect to the small ranges with metallic lines; thus, we decided to use RV estimates from the full spectral range as the final ones for further analysis. The standard deviation of individual RV measurements based on the full spectral range and the standard deviation of mean RV estimates based on the three spectral pieces mentioned above are presented in Fig.~\ref{errors_estimate}. There, we can  see the dependency of presented error bars from the resulting spectral S/N. Also, in the case of spectroscopic binaries (which have been visually identified), the standard deviation of mean RV based on three different spectral regions could significantly increase ($\sigma$ RV > 10 \kms). It is important to mention that in the case of SB2 systems, our RV measurements are potentially unreliable because we used only one stacked template to cross-correlate the spectra with each other. Thus, if a prominent second companion is present in the spectrum, the CCF could give erroneous estimates of RV. Nevertheless, it does not affect our results because the final spectroscopic binary classification for all targets, apart from the RV measurements, is also based on the line profile variability analysis (see the next subsection). 

The next step was to derive the peak-to-peak amplitude of radial velocity variations (RV$_{\mathrm{PP}}$). We defined it as RV$_{\mathrm{PP}}$ = max(|RV$_{i}$ - RV$_{j}$|), where ${i}$ and ${j}$ denote one of the nine epochs. Following \citet{Sana_2012}, \citet{Dunstall_2015} and other accompanying BLOeM papers, each of the RV$_{\mathrm{PP}}$ estimates should satisfy the significance criteria that we have defined according to the next rule: 
|RV$_{i}$ - RV$_{j}$| / $\sqrt{\sigma_i^2+\sigma_j^2}$ > 4, where $\sigma_i$, and $\sigma_j$ are the individual RV uncertainties at epochs ${i}$ or ${j}$. If we detected that the significance criteria had not been satisfied, we examined the quality of the spectra and in case of very low S/N spectra (S/N < 50), we excluded that spectrum from the analysis. The uncertainty associated with the final estimate of RV$_{\mathrm{PP}}$ is computed as an error propagation of $\sigma_i$ and $\sigma_j$. We listed the RV$_{\mathrm{PP}}$ values for all targets in Table \ref{table:all_bsgs}.

\begin{figure}[!ht]
\centering
\includegraphics[width=0.49\textwidth]{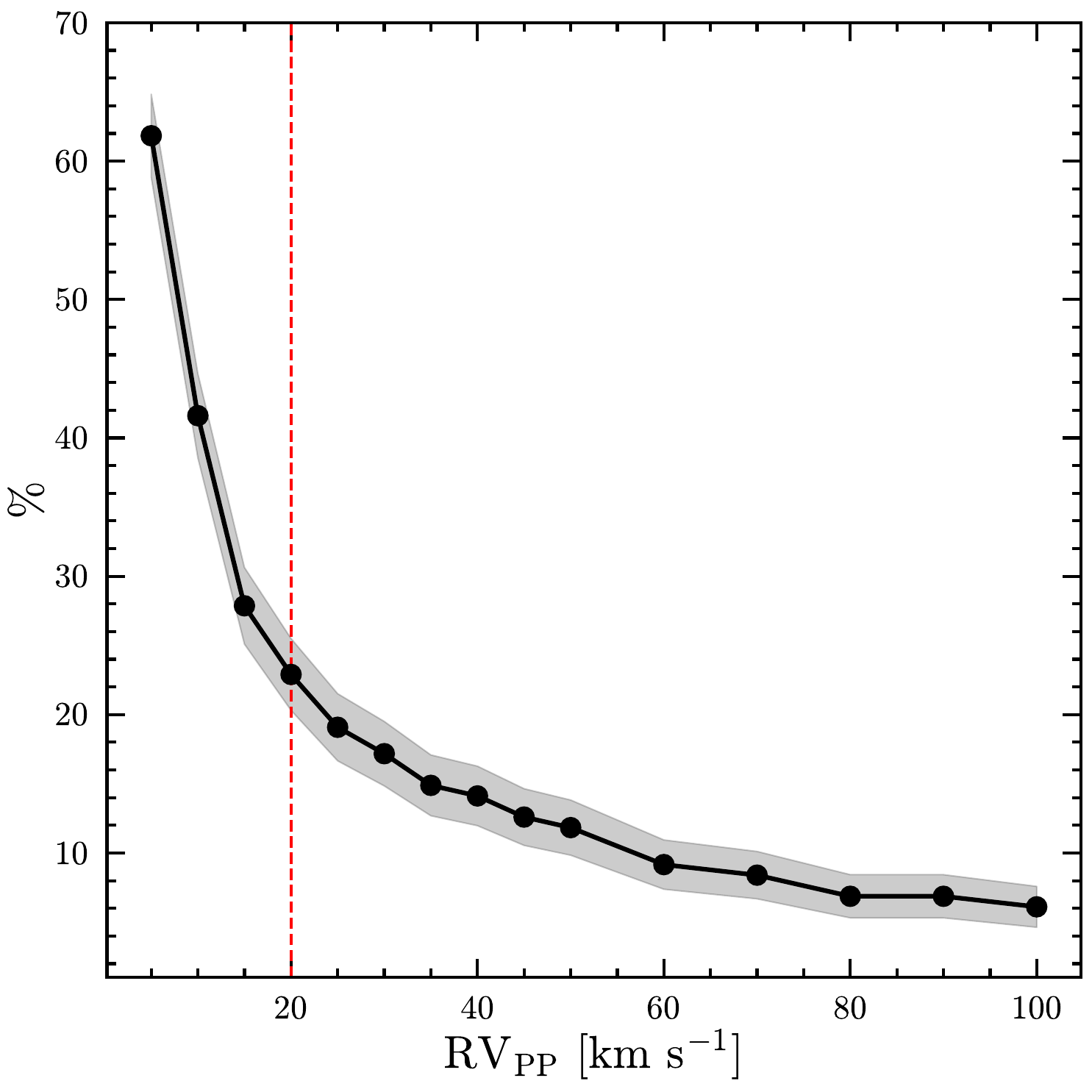}
\caption{Fraction of BSGs with \RVpp larger than a given threshold. The gray shaded area represents the binomial error associated with the corresponding value of the binary fraction.}
\label{figure_rvpp_fraction}
\end{figure}

Figure~\ref{figure_rvpp_fraction} illustrates the fraction of BSGs with RV$_{\mathrm{PP}}$ larger than various RV thresholds. The grey shaded area indicates the binomial error for each calculated fraction of those BSGs in our sample. As a next step, we calculated the spectroscopic binary fraction of BSGs above the given RV variations. 
Following the \citet{Sana_2013} methodology, we considered a star to be a probable binary if it passes the significance criteria and has RV$_{\mathrm{PP}}$ > 20 \kms. This choice of threshold is based on the expected intrinsic RV variability among OB supergiants \citep[see][]{Ritchie_2009,bsgs_iacob}. In this way, as a first-order approximation, we can already estimate the spectroscopic binary fraction of our BSG sample.  However, this approach is very sensitive to the adopted threshold. For example, as can be seen in Fig.~\ref{figure_rvpp_fraction}, by decreasing the threshold from 20 \kms down to 10 \kms, the binary fraction increases from 24$\%$ up to $\sim$43$\%$. At this step, for a correct interpretation of results, it is very important to understand the other sources of RV variations, which could lead to the significant RV$_{\mathrm{PP}}$ values.

\subsection{Line profile variability}
The BSG domain is characterized by significant intrinsic pulsation because at this evolutionary phase, the stars suffer from numerous intrinsic instability mechanisms. For instance, late O-type and B-type stars are known to exhibit RV variations caused by non-radial pulsations (caused by heat-driven p- and g-modes and internal gravity waves) such as those present in slowly pulsating B stars or $\beta$ Cephei stars \citep[see e.g.,][]{Godart_2017,Burssens2020}.

Consequently, pulsations could affect the shape and position of absorption lines which we can interpret as hints of binarity. Such a line profile variability could show RV$_{\mathrm{PP}}$  variations up to 30 \kms~ \citep[see e.g.,][]{bsgs_iacob}.
One example of such intrinsic variability in combination with binary motion can be seen for the system BLOeM 1-103 (left panel in Fig.~\ref{rv_examples}). The binary shows a significant global trend of primary component movement and furthermore, there are fluctuations in the RV variations (third and fourth epoch) caused by the intrinsic variations associated with stellar pulsations. Thus, to avoid false-positive detections of binaries, we decided to make a visual inspection of line profile variability for all available spectra.

We examined how the shapes of line profiles changed within the available epochs and grouped the targets into the following categories, referring to their spectroscopic binary status: (i) apparently single or line profile variable (lpv), (ii) single-lined spectroscopic binary (SB1), (iii) intermediate between SB1 and lpv (i.e. non-prominent spectroscopic binary, lpv/SB1), (iv) double-lined spectroscopic binary (SB2), and (iv) intermediate between SB1 and SB2 (SB1/SB2).  The main criterion we follow to distinguish between these sub-categories is the behavior of line profile variability -- if the wings of the line do not change their position, but the core is moving significantly, we consider it as "lpv." This  corresponds to the spectroscopic binary status of "apparently single" in our formalism. In the case of low S/N, when it is difficult to distinguish movements of the wings, we marked it as "lpv/SB1;" if we spot a clear motion of the entire line, we classified it as "SB1" (we present the examples of line profile variability for each of these types in Fig.~\ref{lpv_examples}). The class of SB2 usually shows a clear change in the morphology of line profiles or if it is unclear we mark it as "SB1/SB2". These visual inspections help to understand the nature of RV variations and, taken together with the  RV$_{\mathrm{PP}}$ measurements, they allow us to describe the sample more comprehensively. 

\begin{figure}[!ht]
\centering
\includegraphics[width=0.49\textwidth]{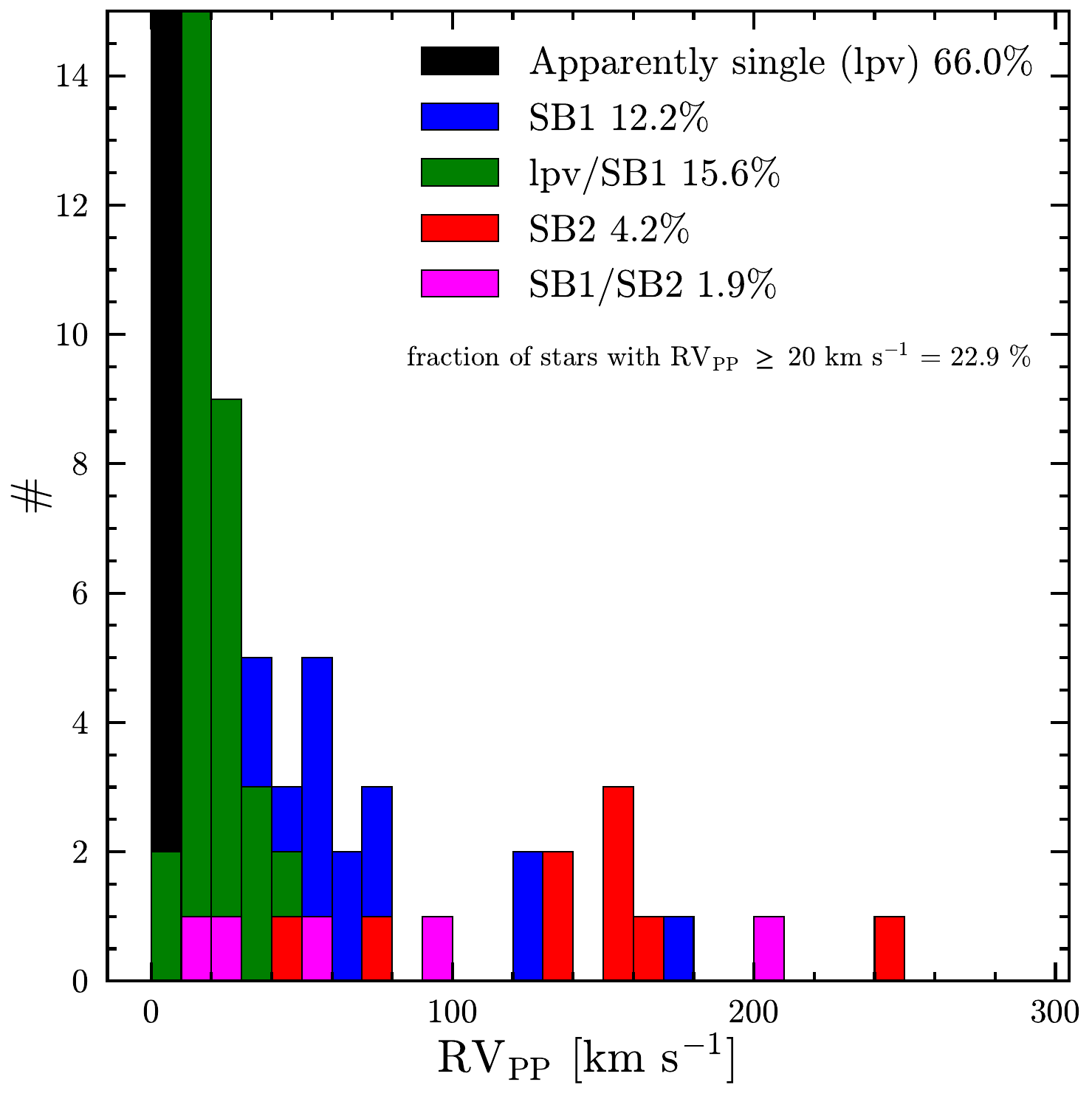}
\caption{\RVpp distribution for the full sample of BSGs, including their spectroscopic binary status. For clarity of the presented data, the ordinate axis has a limit of 15. The number of stars in the first two columns of the histogram is 150 and 24, respectively.}
\label{rvpp_distribution}
\end{figure}

In Fig.~\ref{rvpp_distribution}, we show the resulting RV$_{\mathrm{PP}}$ distribution together with splitting targets by spectroscopic binary status. As we can see there are a few SB1/SB2 systems with a small \RVpp that could be missed without visual inspection of line profile variability. Targets classified as "lpv/SB1" have \RVpp up to 50 \kms; however, we need to admit that such a classification depends a lot on the available number of observations. More available epochs will increase the accuracy of our classification and the "lpv/SB1" class of targets should become less numerous in comparison to what we have in the present work. The overall classification and \RVpp results for each BSG are presented in Table \ref{table:all_bsgs}.

\subsection{Periodogram analysis}
We applied Lomb-Scargle (LS) periodogram analysis
\citep{Lomb_1976,Scargle_1982} to plausibly detect  the most reliable orbital period that would be capable of describing small RV variations. We searched for orbital periods in the range between 0.4 and 500 days. While this range is rather broad, it potentially could help us to detect long-period binary candidates. For the LS analysis, we used its implementation in MINATO code \footnote{https://github.com/jvillasr/MINATO/}. The fitting of only nine available epochs with a sinusoidal function requires a lot of extra checks for the result's reliability. Indeed, with such a cadence of observations, it is easy to fit them with any apparent short periods (i.e., $<$ 2 days). 
In addition, the aforementioned 2 day period serves as a minimum period boundary, which is defined as the Nyquist frequency for a given minimum interval ($\sim$one day) between the observations in the BLOeM campaign.

To avoid any false-positive period finding, we computed the false-alarm probability (FAP) to estimate the probability of having a peak in terms of LS power parameter of a certain height at a given frequency \citep[see the broad definition of this parameter in][]{VanderPlas_2018}. 
The FAP shows the significance of the derived period for the non-periodic signal with respect to the rest of the available solutions. Following \citet{Villase_2021}, we can determine whether the FAP of the corresponding peak is below the 0.1$\%$ of the FAP level and, thus, we can consider the star as SB1 (which means that a given system has a determined orbital period with a probability less than 0.1$\%$ of the FAP). However, as we  point out above, the BSG domain overlaps with the $\beta$ Cephei period regime for pulsations; thus, apart from the prominence of the LS power peak with respect to the FAP level, we consider the absence of similar peaks on the periodogram. While we performed a visual inspection of periodograms, we notice that a lot of targets with a small \RVpp that have been classified as apparently single or "lpv/SB1" have a lot of similar peaks that correspond to the periods of less than 2 days. 
However, those similar peaks on the periodogram never have FAP of less than 1$\%$ FAP level. In this case, we do not consider estimated periods as bona fide; meanwhile, if we detected only one prominent peak with an LS power larger than the power corresponding to the 0.1$\%$ of FAP, we confidently marked it as a reliable period estimate. Notably, we never detected several peaks with such a high LS power; thus, a given conservative FAP threshold gives reliable orbital period estimates. Examples of the most representative periodograms and line-profile variability types for different targets are presented in Fig.~\ref{lpv_examples}.


The estimated periods as a function of \RVpp for all our targets are presented in Fig.~\ref{period_distribution}. We highlighted by colors only those BSGs that have peaks on the periodogram higher than 0.1$\%$ of false-alarm level. It is a conservative selection, however, in this way, we were able to reduce the number of false-positive detections of spectroscopic binaries. In total, we found 41 spectroscopic binaries (SB1, SB2, lpv/SB1, and SB1/SB2) with bona fide periods and two lpvs with small \RVpp variations. The rest of the targets require more observations for their definitive classification. However, we already can conclude that there are a lot of lpvs that could be pulsating variables, especially with a \RVpp < 10 \kms~ and periods less than 2 days. 

The estimated periods were compared with 17 feclipsing binaries \citep[EBs, based on the OGLE-IV data; see][]{Pawlak_2016} which are known in our sample (see the "notes" column in Table \ref{table:all_bsgs}). The vast majority of SB2 systems are EBs with periods of less than 10 days, while most of the SB1 systems are not eclipsing (see Fig.~\ref{period_distribution}). 
The photometric and RV variation periods are identical apart from a progressive discrepancy towards long-period systems BLOeM 4-077 ($P_\mathrm{orb}$ $\sim$49.2 d), BLOeM 5-040 ($P_\mathrm{orb}$ $\sim$128.7 d), and BLOeM 3-073 ($P_\mathrm{orb}$ $\sim$293.7 d). This is expected because nine epochs distributed within three months are not efficient in detecting reliable periods with more than 45 days. Indeed, to constrain successfully the orbital solutions from the light- or radial-velocity curves it is necessary to fit twice the phase-folded data,  to avoid the uncertainties in orbit eccentricity, and so on. 

Other problematic targets (six in total) were the ones for which we detected a period of less than one day based on the RV measurements, however, their photometric light curves show the periods in a range of 1.4 -- 5.1 days. It is explained by the fact that three of these systems are SB2s (BLOeM 7-116, BLOeM 5-075, BLOeM 4-067). Thus, it is necessary to disentangle the spectra first in order to get true RV estimates of the companions and consequently reliable period estimates. Another three systems are SB1s, one of which deserves particular attention. The BLOeM 4-068 has a significant \RVpp variation (74.80 \kms) and based on our periodogram analysis, it exhibits $P_\mathrm{orb}$ $\sim$95.24 days, however, based on OGLE data the period is 1.39 days. 
The light curve shows the tiny eclipses ($\sim$0.04 mag) and such a short observed period as we would expect for the contact binary configuration of two B-type MS stars. On top of that, there is a small amplitude signal at $\sim$183 days, which could be associated with a double of $P_\mathrm{orb}$ we detected from RV measurements. Thus, we can assume this system is a triple, with two MS stars on the inner orbit and BSG on the outer one.



To conclude, given the limited number of epochs and time cadence of the observations, the most reliable range of the estimated periods of RV variations in our analysis should be between 2 and $\sim$50 days. 
These period boundaries are determined by cadence and time span of presently available BLOeM observations. If we aim to fit the phase-folded radial velocity curve twice, the maximum period boundary is $\sim$25 days, while we can extend this boundary to the observed range of observations $\sim$100 days if preliminary period estimates are required. 
In the case of systems with longer periods ($P_\mathrm{orb}$ > 100 days), we must increase the cadence and time span of observations because such systems can be eccentric; thus, we cannot use the LS approach where we can fit only the circular orbits. 


\subsection{Exploration of the minimum boundaries for the orbital period of spectroscopic binaries}

The detection of a few extremely short-period ($P_\mathrm{orb}<$ 2 days) SB1 and SB2 systems raises the question of whether such BSG systems could physically exist. To check this, we computed the Roche-lobe filling factor for all sample of BSGs, defined as the ratio of the radius of the assumed primary component to its Roche-lobe radius (R$_\mathrm{RL}$).
We computed it for all targets because such a test can indicate which estimates of minimum orbital periods are not physically realistic. 
We used the \citet{Eggleton_1983} fitting formula to compute the R$_\mathrm{RL}$ for all spectroscopic binaries by assuming the constant mass ratio $q=M_2/M_1=$ 0.55 and their primary components' evolutionary masses and radii. The preliminary estimates of evolutionary masses have been derived using the BONNSAI code \citep{Schneider_2014}, more details will be published in the accompanying paper (Bestenlehner et al. in prep.). We did not see any significant difference in the resulting Roche-lobe filling factor (R$_\mathrm{1}$/R$_\mathrm{RL}$) after varying the values of the mass ratio. The estimated  R$_\mathrm{1}$/R$_\mathrm{RL}$ as a function of the orbital periods is presented in Fig.~\ref{period_distribution} (bottom panel). Notably, all binaries with $P_\mathrm{orb}>$ $\sim$3 days have the Roche-lobe filling factor less than one. It could serve as an additional constraint to the minimum boundary of the orbital periods because such systems cannot physically exist; namely, the systems with a Roche-lobe radius of primary component less than its actual radius. However, taking into account the uncertainties in the radii estimates we can adopt the maximum boundary of a Roche-lobe filling factor of 1.5 that corresponds to our minimum boundary of detectable bona fide orbital periods equal to 2 days (which agrees with the existence of EB in such period regimes).
In addition, we can see in this diagram that all spectroscopic binaries with orbital periods up to 7 days are potentially interacting systems, taking into account their relatively high Roche-lobe filling factor (R$_\mathrm{1}$/R$_\mathrm{RL}$>0.8). This is evidence that this group of targets is a transition evolutionary stage of systems with an ongoing mass transfer. Nevertheless, observationally they look like BSGs, however, they do not represent the single-star evolution path of the targets that may also be in this BSG domain.

\subsection{BLOeM survey capability and intrinsic binary fraction}

We used the Monte Carlo population-synthesis method presented in \cite{Sana_2013} to quantify the ability of the BLOeM campaign to detect BSG binaries of different orbital periods, mass ratios, and eccentricities. In addition, we adopted an improved version of the method that includes uncertainties on the underlying distributions \citep[according to][]{Banyard_2022}. Specifically, we simulated 10\,000 observing campaigns of 262 BSGs, adopting the temporal sampling and RV uncertainties.  We adopted power law representations for the distributions of orbital period $P_\mathrm{orb}$ (within the range: 1 -- 10$^{3.5}$), mass ratio (0.1 -- 1.0), eccentricity ($e$, 0 -- 0.9) by assuming random spatial orientations of the systems. The primary mass ($M_{1}$) ranges from 16 to 60 \Msun following the \citet{Salpeter_1955} initial mass function. In addition, we apply a circularisation criteria of the shortest-period systems.

In Fig.~\ref{bias_diagrams}, we present the outcome of detection probabilities of binaries in terms of the mentioned orbital parameters.
The resulting detection probability of the survey within the full range of periods log(P) = [0.0 -- 3.5] is 0.66 $\pm$ 0.06 and within $P_\mathrm{orb}$ < 365 d: 0.89 $\pm$ 0.03. 
Taking into account the derived detection probabilities and assuming the threshold of \RVpp equal to 20 \kms~ for our sample of BSGs, the corrected (intrinsic) binary fraction appeared to be 40 $\pm$ $4$$\%$. 

\begin{figure}[!ht]
\centering
\includegraphics[width=0.49\textwidth]{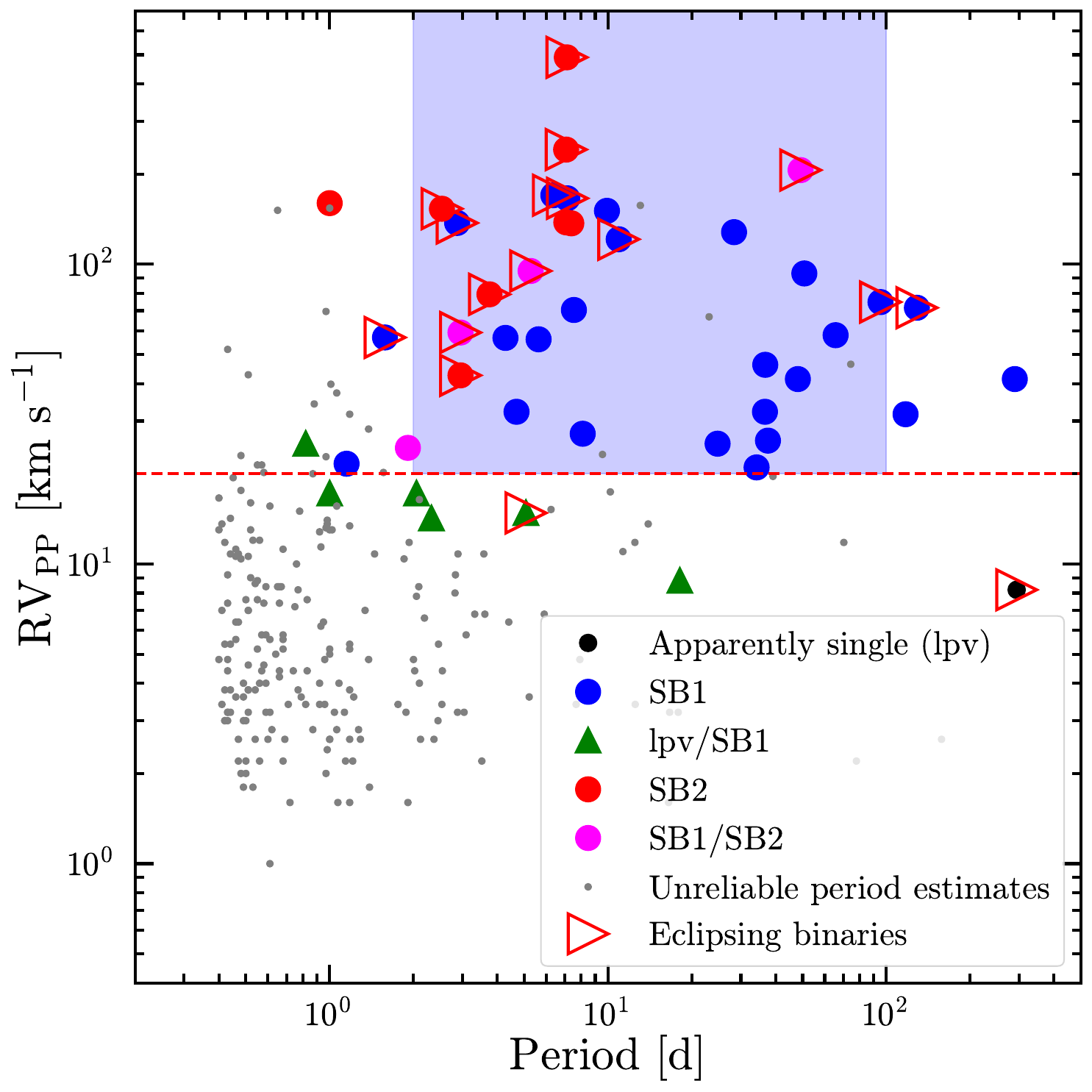}
\includegraphics[width=0.49\textwidth]{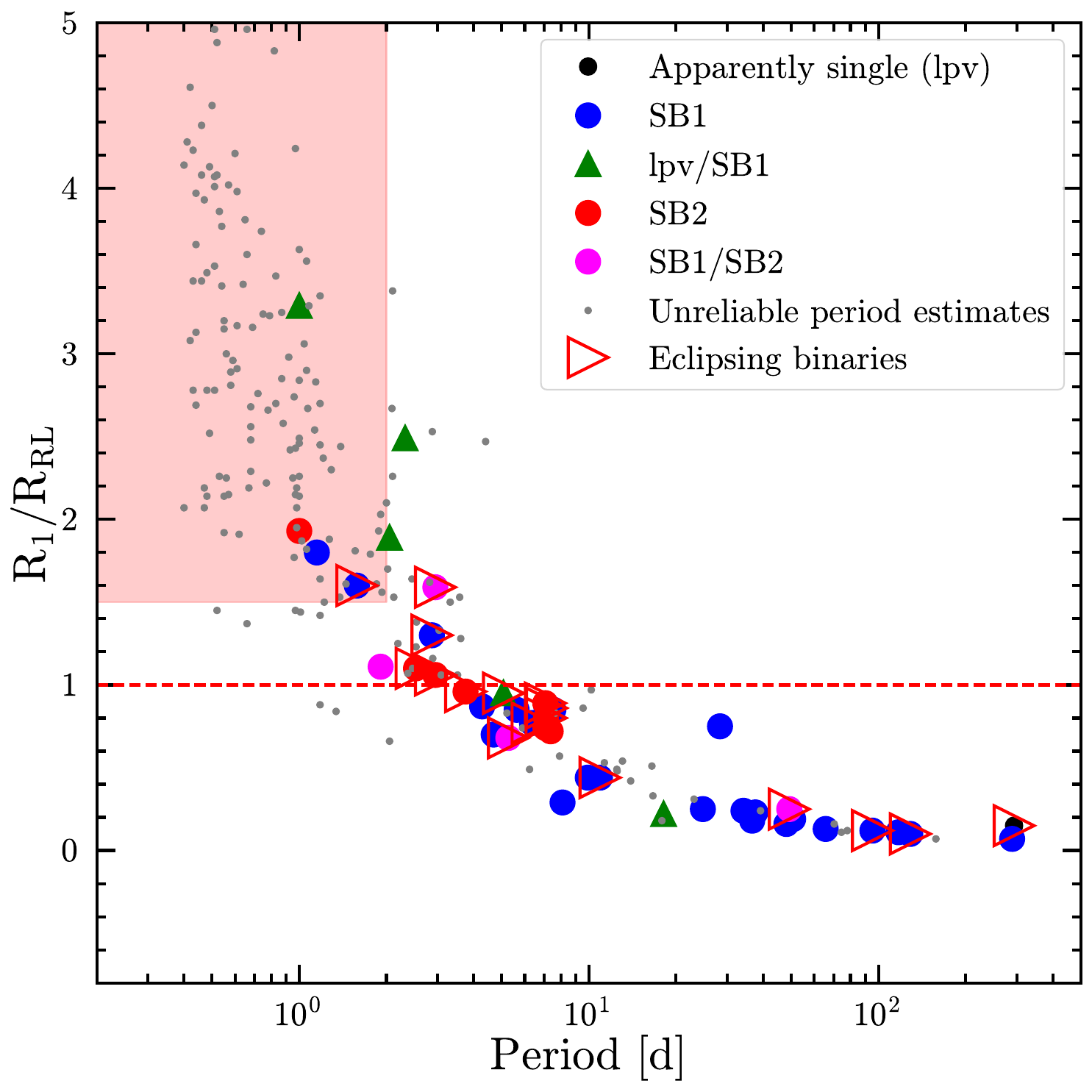}
\caption{Estimated periods of radial velocity variability as a function of \RVpp (top panel) and the Roche-lobe filling factor (bottom panel) for the entire sample of BSGs together with their spectroscopic binary status. We colored only the targets with the FAP below the 0.1$\%$ of FAP level. All detected eclipsing binaries are marked accordingly. The blue-shaded area represents the region where orbital periods and spectroscopic binary classification for all BSGs are well-constrained given the limitations of available data. The red-shaded area (bottom panel) indicates the non-bonafide region in terms of estimated orbital periods (<2 days) and Roche-lobe filling factor (R$_\mathrm{1}$/R$_\mathrm{RL}$>1.5).}
\label{period_distribution}
\end{figure}

\begin{figure}[!ht]
\centering
\includegraphics[width=0.49\textwidth]{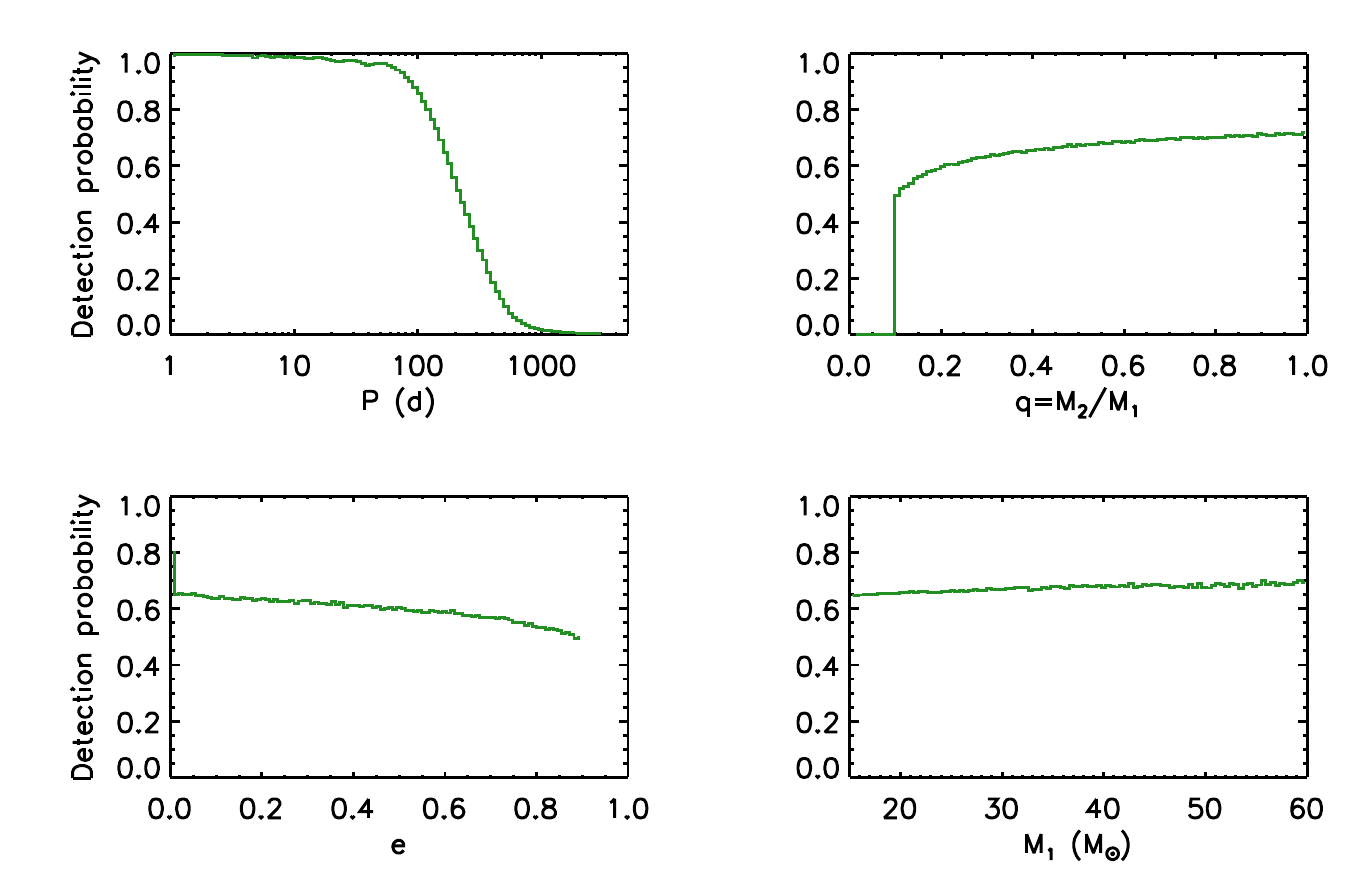}
\caption{Binary detection probability of the BLOeM survey for BSGs. It shows that the given cadence of observations has excellent detection capability for periods shorter than $\sim$100 days, which agrees with our periodogram analysis.}
\label{bias_diagrams}
\end{figure}

Another confirmation of our simulated detection probability is that, indeed, we were able to find binaries with $P_\mathrm{orb}$ $\lesssim$ 100 days (see Fig.~\ref{period_distribution});  however, even with the nine available epochs distributed over three months, it is possible to have a zero-order approximation regarding the periods of long-period binary systems ($P_\mathrm{orb}$ > 100 d), as we detected a few spectroscopic binaries with periods $\sim$100 -- 150 days.

\begin{figure*}[!ht]
\centering
\includegraphics[width=0.50\textwidth]{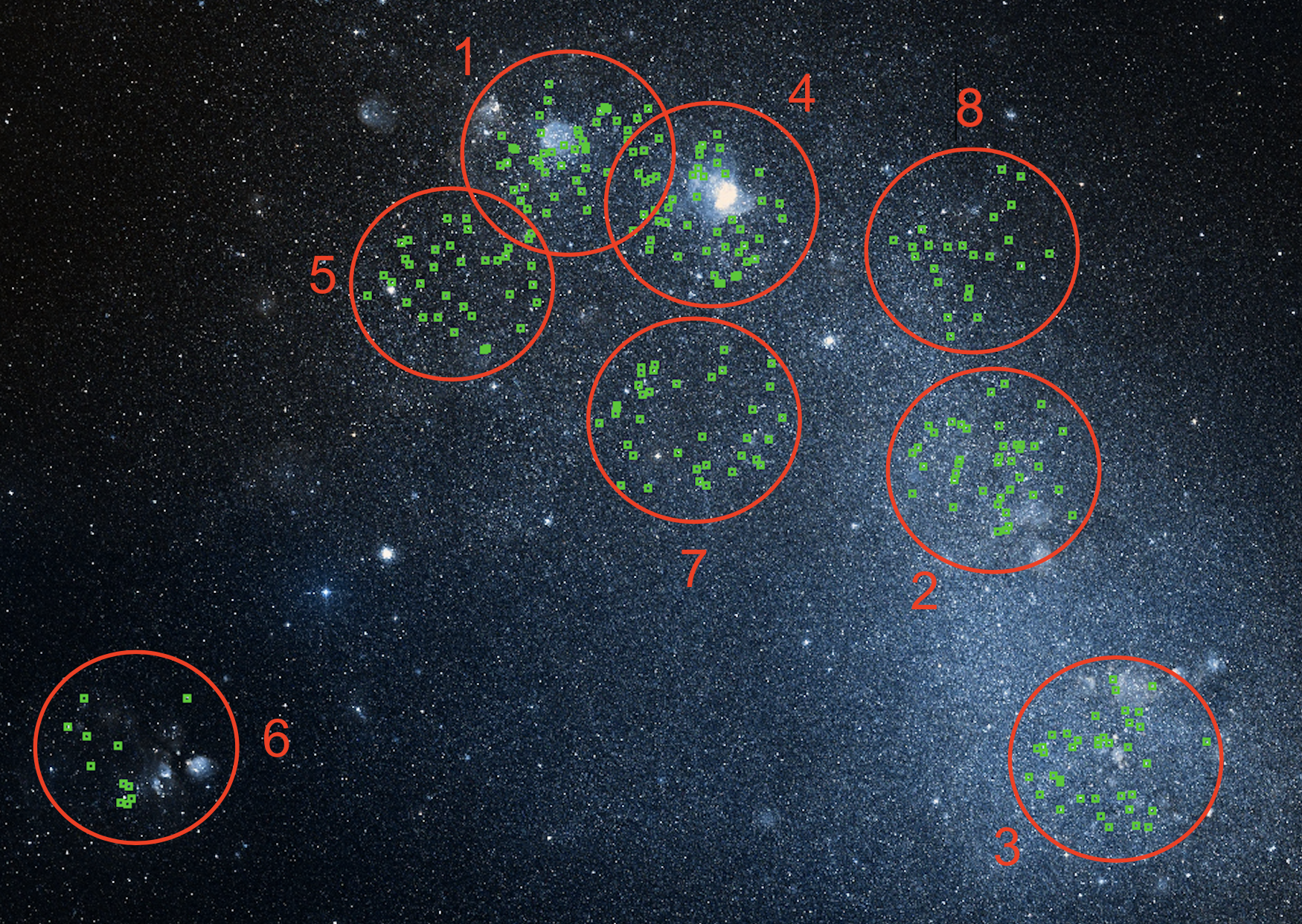}
\includegraphics[width=0.49\textwidth]{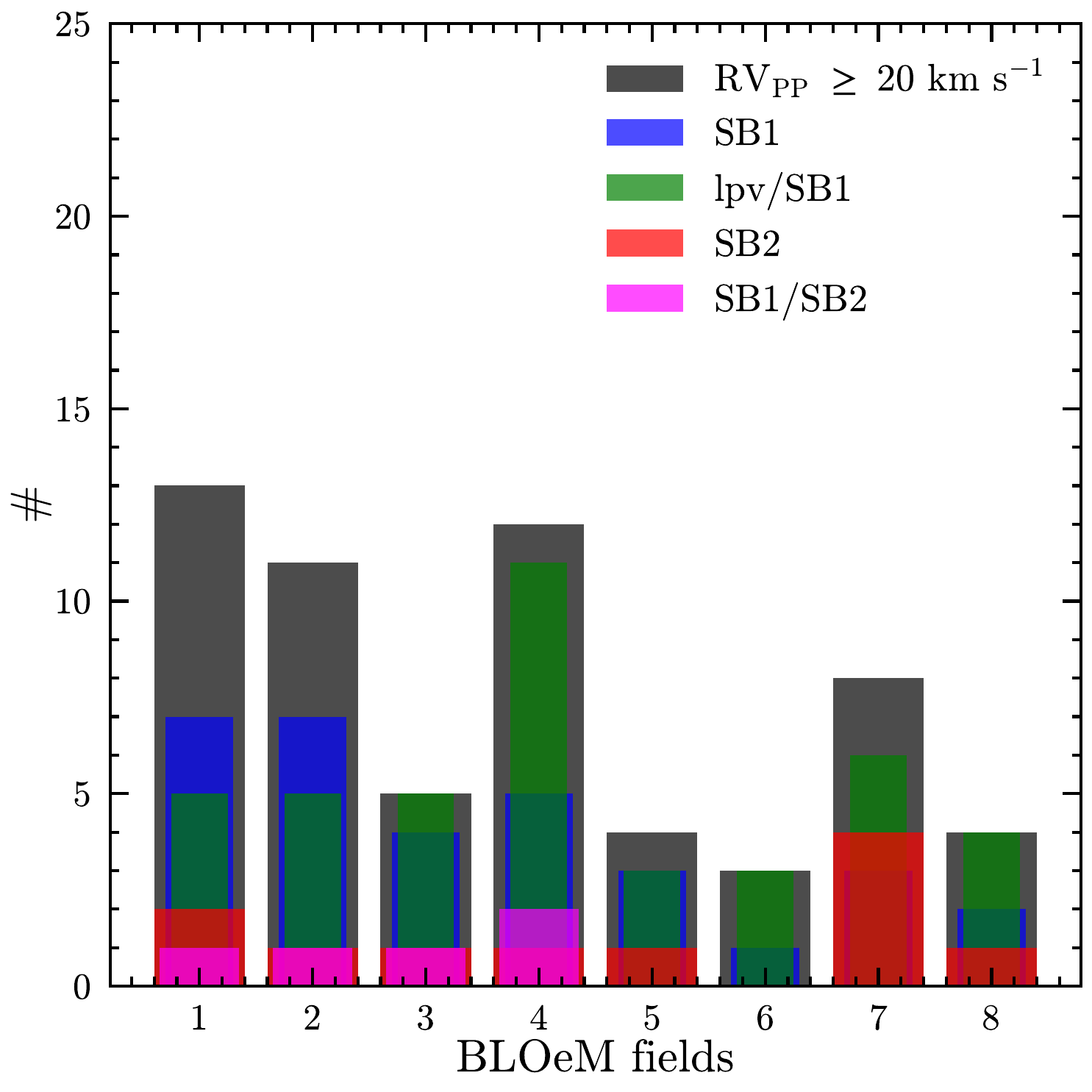}
\caption{Left panel: Spatial distribution of investigated BSG sample in each BLOeM field superposed on Digitized Sky Survey image of the SMC (STScI/NASA, Colored \& Healpixed by CDS). Right panel: Number of BSGs with various binary status for each BLOeM field. The number of targets with \RVpp > 20 \kms~is not connected with the number of targets visually classified as spectroscopic binaries.} 
\label{field_distribution}
\end{figure*}


\begin{figure*}[!ht]
\centering
\includegraphics[width=0.49\textwidth]{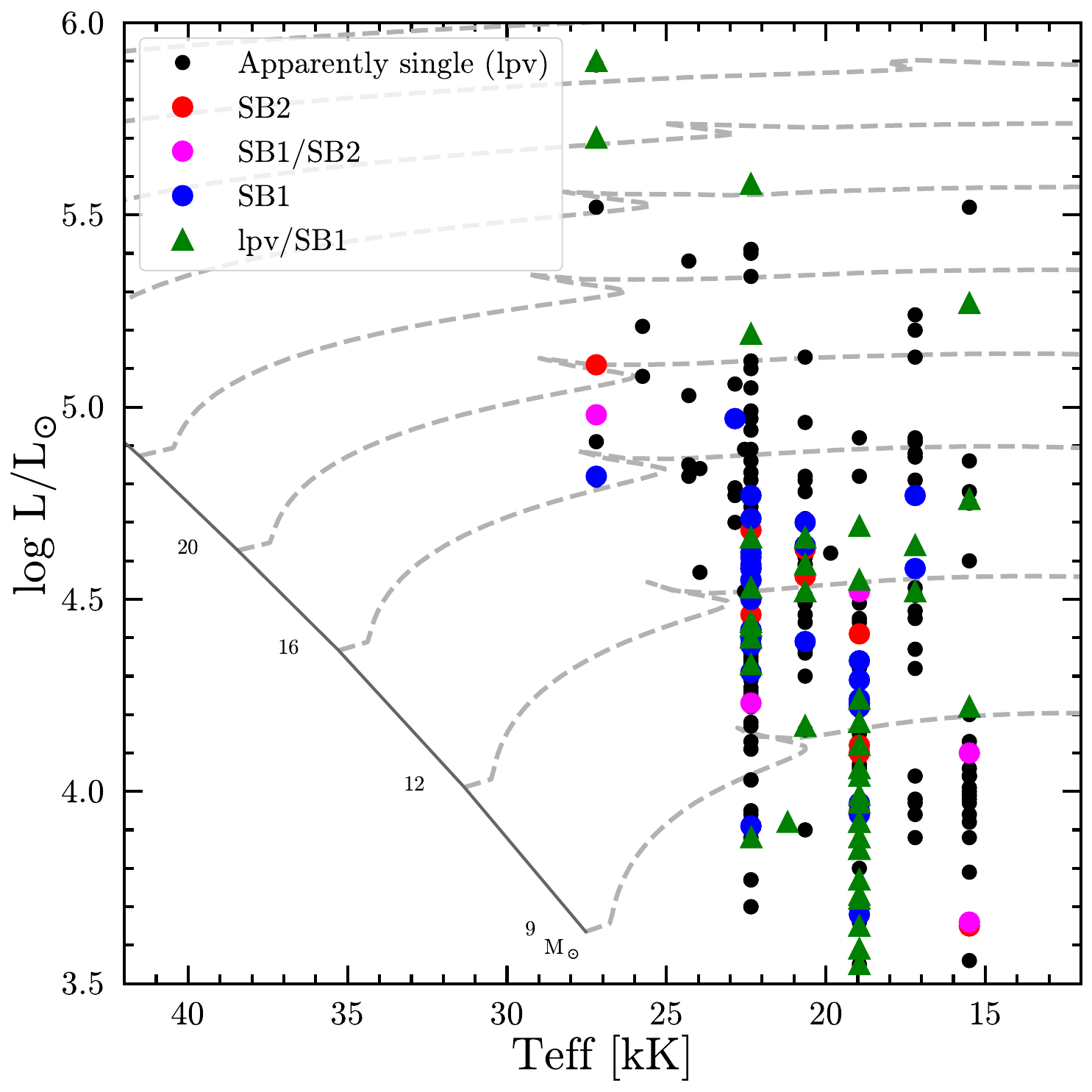}
\includegraphics[width=0.49\textwidth]{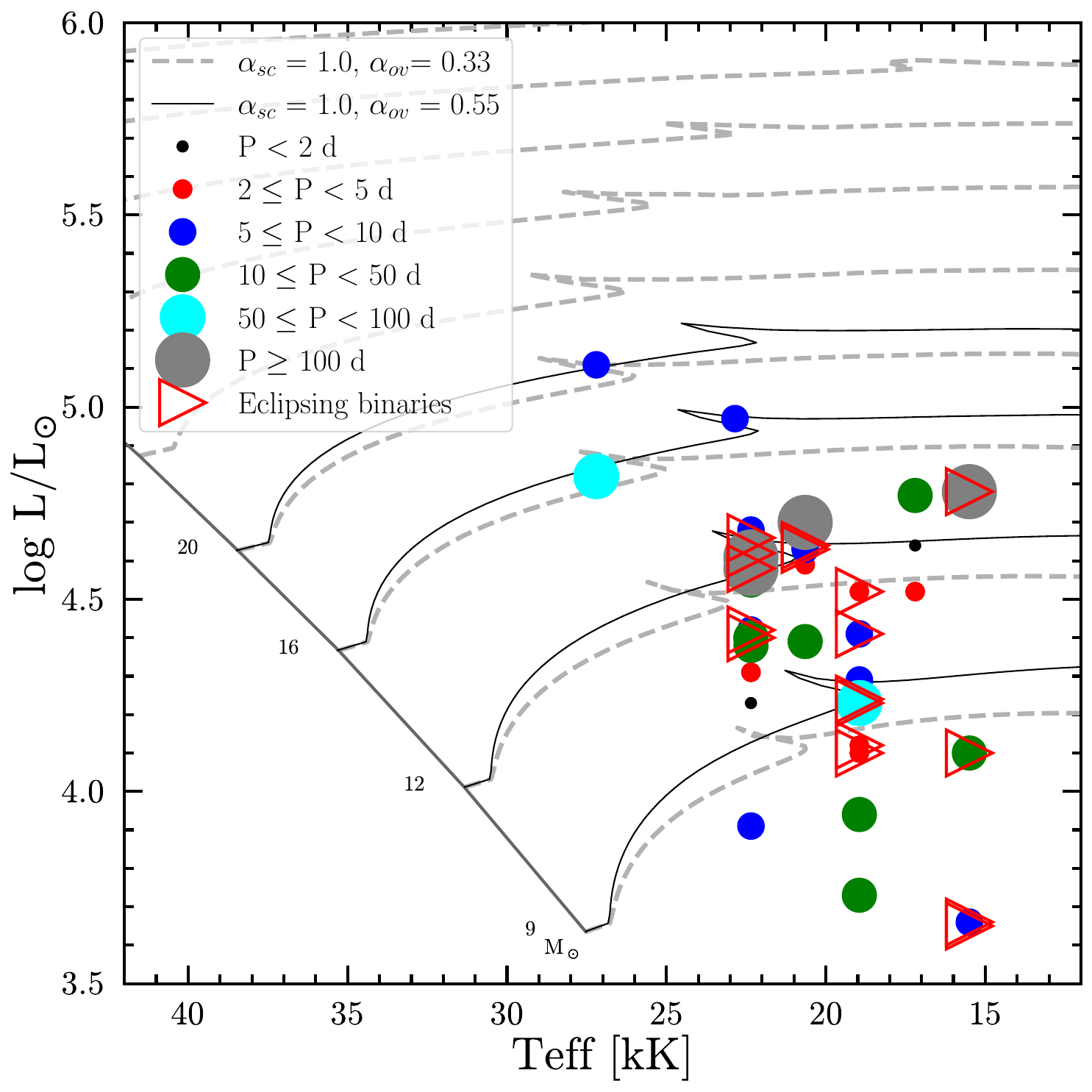}
\caption{Position of the whole studied sample of BSGs (left panel) on the Hertzsprung–Russell diagram together with non-rotating single star evolutionary tracks for the SMC metallicity \citep[][]{Schootemeijer_2019}. The given tracks were calculated by adopting the efficiency of semiconvection ($\alpha_{sc}$) equal to 1 and the convective core step overshooting efficiency ($\alpha_{ov}$) equal to 0.33. The right panel represents the position of only "bona fide" spectroscopic binary BSGs for which prominent periods of RV variations were constrained. In addition, four evolutionary tracks with adopted $\alpha_{ov}$ = 0.55 were added for reference (black solid lines).} 
\label{hr_diagram}
\end{figure*}

\begin{figure*}[!ht]
\centering
\includegraphics[width=0.49\textwidth]{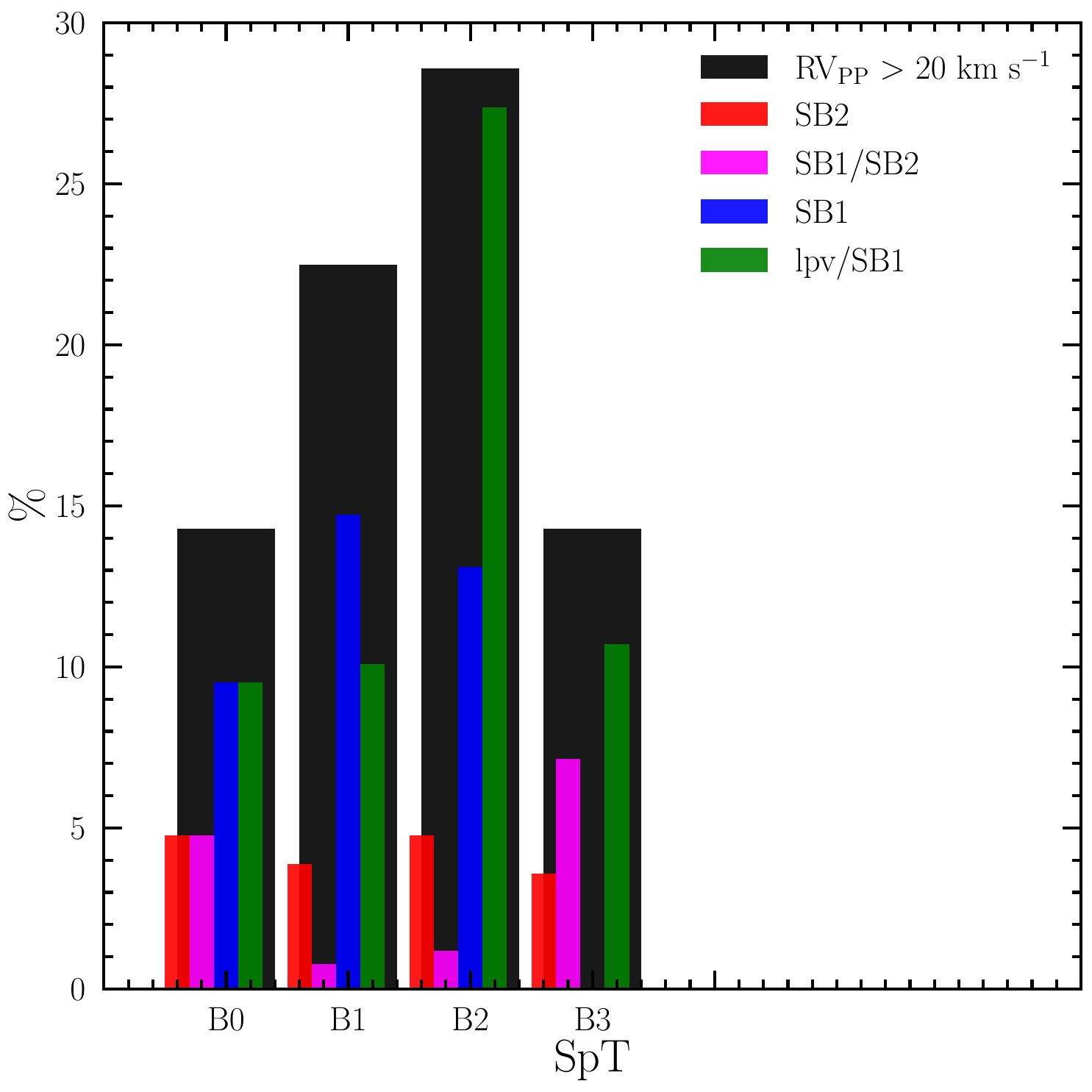}
\includegraphics[width=0.49\textwidth]{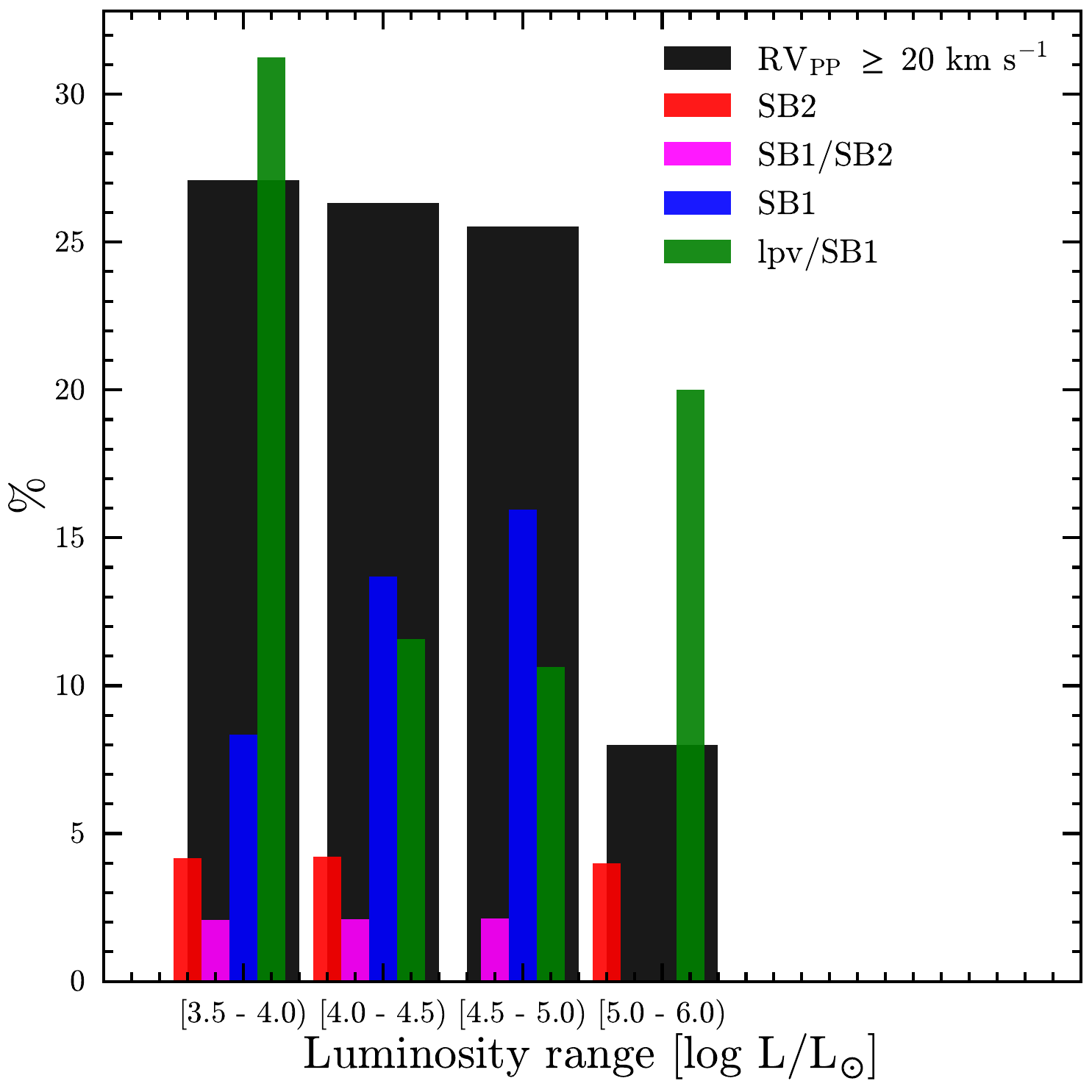}
\caption{Fraction of spectroscopic binaries among the program sample of BSGs calculated for each spectral sub-type (left panel) and luminosity range (right panel).}
\label{spt_teff_distribution}
\end{figure*}

\begin{figure}[!ht]
\centering
\includegraphics[width=0.49\textwidth]{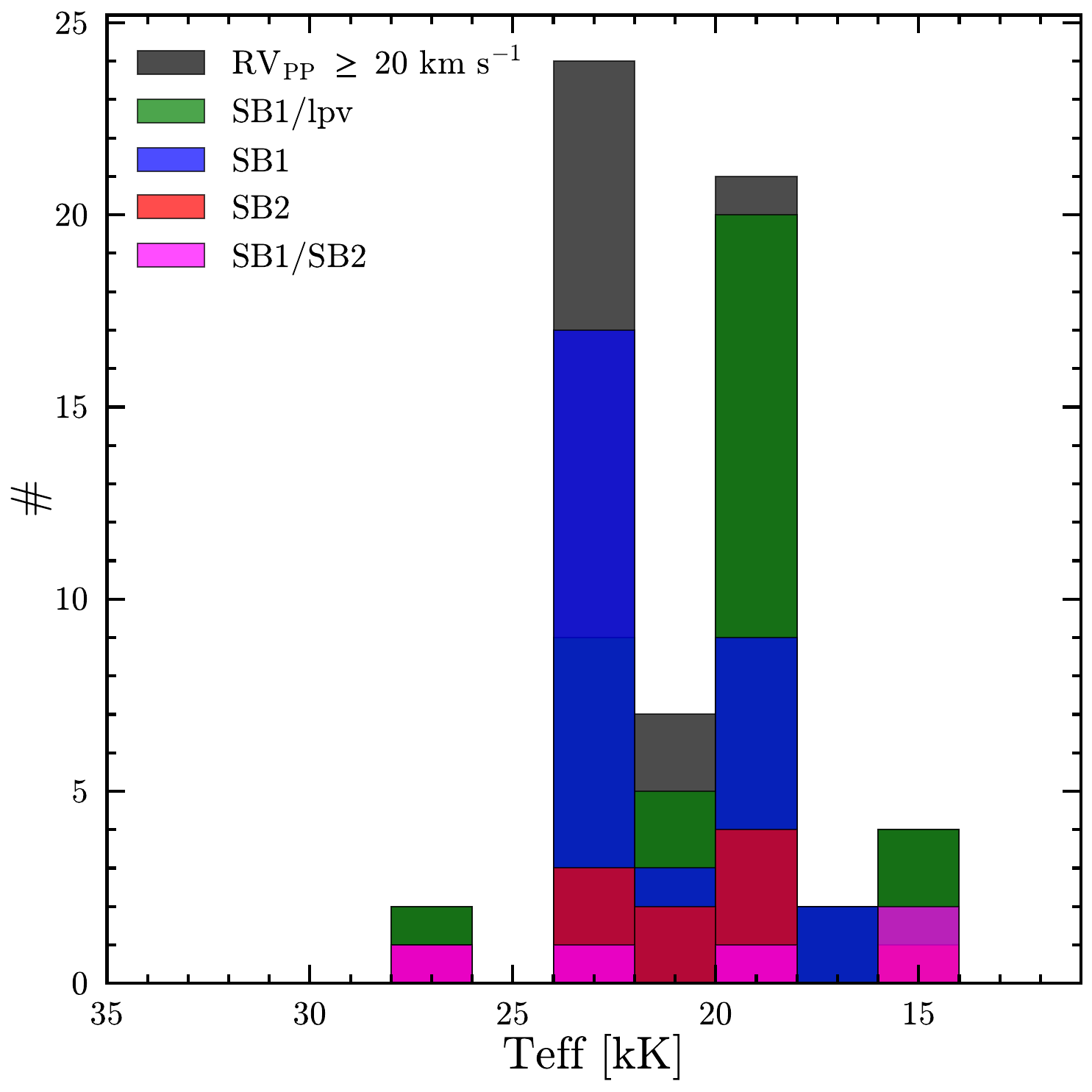}
\caption{Distribution of studied BSGs grouped by their preliminary estimated effective temperatures.}
\label{teff_distribution}
\end{figure}

\section{Discussion}

\subsection{Multiplicity of BSGs}

Using 20 \kms~as a threshold for the peak-to-peak RVs, we obtained an observed spectroscopic binary fraction of 23 $\pm$ 3\%. However, as we already pointed out, using a simple threshold for such stars that could have a significant intrinsic variability (such as pulsations) is risky in terms of false-positive identification of binaries. In this regard, we included line profile variability as a final criterion to consider the star as a spectroscopic binary.  
Interestingly, these two approaches complement each other, it is well seen in the distribution of \RVpp as a function of estimated orbital periods (see the top panel in Fig.~\ref{period_distribution}). If we analyze the correlation of \RVpp values versus binary status we see that above \RVpp of 20 \kms~ we have only spectroscopic binaries and one lpv/SB1; meanwhile, in the range of  10 < \RVpp < 20 \kms~ we can see a few lpv/SB1s and one EB that we would miss by using a simple \RVpp threshold cut.
The fraction of spectroscopic binaries in our sample including "lpv/SB1s" is 34 $\pm$ 3\%, which is significantly larger than simply considering a threshold of 20\kms.


It is worth mentioning that the information about EB status also helps to confirm spectroscopic binary status, but not in the case of long-period systems (>100 days). Indeed, we classified one BSG (BLOeM\_3-073) with \RVpp = 8.2 \kms~ as apparently single; however, but this is, in fact, an EB with a $P_\mathrm{orb}$ $\sim$293 days \citep[according to][]{Pawlak_2016}. This outcome is expected because nine epochs spread over 90 days are insufficient for detecting such a long-period system. 
Apart from this target, all EBs we classified as spectroscopic binaries (see Fig.~\ref{period_distribution}).  

In Fig.~\ref{field_distribution}, we present the spatial distribution of studied BSGs together with a number of spectroscopic binaries grouped by the  BLOeM observational fields. There is no uniform distribution of binary BSGs across the different fields; notably, there are just a few BSGs in the field number 6, that could be connected with a small number of star-forming regions in that area of SMC.

None of  the BSGs in our sample is detected in X-rays despite good coverage of the SMC galaxy by X-ray surveys \citep[the list of all bright X-ray sources among the BLOeM targets is presented in][]{bloem_i}. This implies that none of the stars in our sample represent a persistent high-mass X-ray binary (HMXBs) containing a compact object embedded in the wind of the donor BSG. Interestingly, in the Galaxy, BSG stars are the most common type of donor stars in wind-fed HMXBs \citep[][]{Silvia_2017}.

The overall determined spectroscopic binary fraction of BSGs in the SMC is similar to those detected in the LMC and in the Galaxy. \citet{Dunstall_2015} analyzed multiplicity properties of 408 B-type stars in the LMC and reported the observed binary fraction among non-supergiants to be  25 $\pm$ 2\% and  23 $\pm$ 6\% for supergiants. In the Galaxy, \citet{burgos_2025} found an average $\sim$27\% of SB1 systems within a large statistical sample of B0-B6 supergiants. Such an independence of BSG binary fraction with respect to the metallicity is in agreement with the binary fraction of early B-type eclipsing systems in the Galaxy, LMC, and SMC \citep[within the $P_\mathrm{orb}$ < 20 d, see][]{Moe_2013}. As \citet{Moe_2019} argued, the explanation for this could be that significant accretion rates in the massive star domain are so high that the massive protostellar disks will always become gravitationally unstable and fragment independent of the metallicity environment (which is not the case in the solar-type star domain). Furthermore, in the early B-type giants and dwarf BLOeMs' sample, it has been found that binary fraction to some extent is dependent on metallicity \citep[for details, see][]{Villasenor_2025}.  
In the next subsection, we discuss the binary fraction as a function of SpT and luminosity.


\subsection{The BSGs landscape on the HRD}

Having obtained the binary properties for such a large number of BSGs along with preliminary estimates of their physical parameters, we have been  able to perform the most intriguing part of our study -- to place them on the HRD. 
In Fig.~\ref{hr_diagram} (left panel), we present the location of all BLOeM BSGs grouped by their spectroscopic binary status. Looking at the HRD, we emphasize the absence of SB1 or SB2 systems above 20\Msun evolutionary track after the terminal-age main sequence (TAMS). Notably, the three most luminous BSGs have significant line profile variability (\RVpp up to $\sim$30~\kms) and, apart from the RV shifts, they have variations in the line-profile depths. The periodogram analysis of RV measurements does not show any prominent frequencies; thus, such line-profile variations could be associated with a strong stellar wind or stochastic pulsations \citep[e.g., see][]{Bowman_2020,Ma_2024ApJ}. To confirm this, it is necessary to study the behavior of the H$\alpha$ region, which is absent in the present spectroscopic data. 

As we estimated the orbital periods for 41 spectroscopic binaries, we can investigate the period distribution of these binaries in terms of their location on the HRD. In Fig.~\ref{hr_diagram} (right panel), we placed only those binaries with reliably derived periods (FAP < 0.1\%) and grouped them according to the different period ranges. BSGs are uniformly distributed in terms of period ranges, apart from a small group of extremely long-period systems ($P_\mathrm{orb}$ > 100 d).
We note that there are no known EBs among the most luminous BSGs. All of our targets are relatively bright sources; thus, we argue that the existing list of EBs in our sample is complete (especially taking into account the sensitivity of the OGLE survey).  

In Fig.~\ref{spt_teff_distribution}, we explore the spectroscopic binary fraction of BSGs in each sub-spectral type (left panel) and in different luminosity ranges (right panel). The histograms present the number of different types of spectroscopic binaries with respect to the total number of BSGs in these specific sub-domains. 
Notably, within the luminosity range of 3.5 < log($L/L_\odot$) <  5.0, the spectroscopic binary fraction
is constant ($\sim$25\%), while in the most luminous domain it significantly drops. This split in binary fractions in terms of luminosity could indicate a different population of BSGs that have different origins, as well as possibly a change in the underlying physics (properties of the winds, etc.).
Indeed, in the absence of spectroscopic binaries, the most luminous BSGs can be interpreted 
as potential merger products, while the predominance of binary systems among the least luminous targets would suggest that they are still on the MS phase. However, taking into account that we analysed nine epochs, it is not possible to detect possible long-period binaries among the most luminous BSGs; thus, the complete BLOeM survey with 25 epochs will extend the binary detection probability to much longer periods.



In addition, if we look at the distribution of spectroscopic binaries for a given spectral type (see the left panel in Fig.~\ref{spt_teff_distribution}), we can notice that the B2 domain contains the largest number of spectroscopic binaries (especially "lpv/SB1") with respect to the rest of the spectral types in our BSG sample. In total, there are up to 50$\%$ of potential spectroscopic binaries (apart from the evolutionary effect), which can also be explained by their relatively low luminosity; thus, some of these targets could belong to the B-dwarf sample, where the observed multiplicity fraction is significant \citep[$\sim$50$\%$, see the accompanying BLOeM paper][]{Villasenor_2025}. Also, some of these least luminous BSGs could be $\beta$ Cephei pulsators, as these tend to be more commonly found among late-O and early-B spectral types \citep{Burssens2020}; thus, this could explain a given overpopulation of lpv/SB1 targets.

Another interesting point is that most short-period spectroscopic binaries are located among the least luminous BSGs, with temperatures hotter than 18 kK \citep[it was also found in the LMC, see][]{McEvoy_2015}. The amount of binaries significantly drops below 18 kK or B2 spectral type (see the distribution of BSGs grouped by $T_{\rm eff}$ in Fig.~\ref{teff_distribution}). The presented distribution is not uniform, there is a clear drop in the number of stars at $T_{\rm eff}$ $\sim$22kK, which can be explained by uncertainties in SpT - $T_{\rm eff}$ calibration; thus, it represents the artificial step in the SpT classification (i.e., B1, B1.5, and B2). However, we do not expect a big difference between the values of physical parameters derived by using photometry and spectroscopy. In our sample, we have seven bright BSGs that have been studied extensively by using ultraviolet mid-res spectroscopy \citep[ULLYSES survey, see][]{Bernini-Peron_2024}. The discrepancy in $T_{\rm eff}$ is within 3 kK and in log($L/L_\odot$) does not exceed 0.1. 

Another reason for non-uniform $T_{\rm eff}$ distribution could be the selection bias of BSGs with the luminosity class of II, indeed, in this domain there is a significant overlap with the B-giants BLOeM sample \citep[i.e., luminosity class of  III, see][]{Villasenor_2025}, which could bias the overall number of BSGs in a given SpT and LC.  Nevertheless, there is a significant drop in binary fraction below 18 kK and even though the present physical parameters of BSGs are not well defined yet, we can argue that it is a real drop in the binary fraction of BSGs at a given effective temperature and spectral type. Moreover, RV variations of late B-type supergiants are significantly smaller \citep[less than 10 ~\kms~ for SpT B5 and later; see][]{Patrick_2025} than the ones studied in the present work, which confirms the absence of short-period supergiants at SpT later than B3. Nevertheless, the binary fraction among B3 BSGs is $\sim$20\%, which agrees with the observed spectroscopic binary fraction of more evolved late-B, A-, and F-type supergiants ($\sim$20\% for the B5-9s SpT domain and significantly smaller binary fraction for the  A- and F-type supergiants). 
It could indicate that at this SpT domain (B2) and $T_{\rm eff}$ regime, we observe the end of MS for a studied sample of BSGs; actually, it will also be interesting to see the behavior of $v$\,sin\,$i$ for the same targets. If there are no fast-rotating ($v$\,sin\,$i$ < 100 ~\kms) BSGs with a SpT later than B2, it can also support the idea of the  MS boundary at this evolutionary phase \citep[as it has been shown in][]{Vink_2010}. Notably, the BLOeM survey has been designed in a way to observe as many as possible bright massive stars in the SMC, while also aiming to achieve a homogeneous sampling across the G-band from 10 to 16.5 mag. The overall sample completeness for the stars with $8 \lesssim M_{\rm ini}/M_\odot \lesssim 14$ (i.e. those born as B-type stars) is $\sim$20\% with respect to the {\it Gaia} catalogue within the range of 13 -- 15 mag in the G-band \citep[see Fig. 2 in][]{bloem_i}. Thus, the given sample of BSGs and the other sub-samples of the BLOeM survey do not have a significant sample selection bias in terms of the magnitude and colour of targets that would affect the interpretation of our results because of the incompleteness of the samples.

Finally, to fit the SMC evolutionary tracks to the position of studied BSGs on the HRD (see Fig. \ref{hr_diagram}), the MS phase should extend to lower temperatures to describe such a large number of binaries in a given $T_{\rm eff}$ and log($L/L_\odot$) regime. Actually, such a significant number of binary BSGs beyond the theoretical TAMS phase is aligned with the results from the LMC, also finding a significant number of binary BSGs beyond the TAMS \citep[see][]{McEvoy_2015}. However, the location of the modeled TAMS depends on certain assumptions and initial parameters \citep[e.g., overshooting efficiency, initial stellar rotation; see][]{Martins_2013}; thus, such observational studies of a large number of BSGs could help to  empirically constrain the position of the TAMS \citep[see also,][for the case of Galactic stars]{burgos_2025}. Indeed, one of the ways to extend the modeled MS phase to that regime is to increase the adopted core overshooting efficiency. We present two sets of tracks, with different overshooting parameter values, in the right panel of Fig.~\ref{hr_diagram}. As we can see, the TAMS of tracks with the adopted overshooting equal to 0.55 could reach most of the detected binary BSGs.

\section{Summary and conclusions}

In this work, we studied the multiplicity properties of a large sample of early B-type supergiants in the SMC within the BLOeM survey. In total, we analyzed the multiplicity properties of 262 BSGs, for which the first nine epochs out of 25 have been available within the BLOeM observational campaign. From the presented analysis of a large sample of BSG in the SMC within the early-BLOeM observational campaign, we can draw the following key conclusions:

\begin{itemize}
    \item The observed spectroscopic binary fraction of BSGs with \RVpp > 20~\kms~is 23 $\pm$ 3\%.  As luminous BSGs could have a significant intrinsic line profile variability (caused by coherent and/or stochastic pulsations) due to strong stellar winds, we made a separate status for spectroscopic binary candidates (i.e. "lpv/sb1"). We added to the same category possible long-period binary systems for which a given range and cadence of observations is not efficient to constraint binary status; however, based on the periodogram analysis, there is a hint that those targets could be binary. By taking into account this non-bonafide "lpv/sb1" class of targets, the spectroscopic binary fraction for our sample is increased to 34 $\pm$ 3\%.

    \item Analysis of the distribution of identified spectroscopic binaries across the different SpTs and luminosity ranges reveals a significant drop of binary fraction after B2 SpT (that corresponds to $T_{\rm eff}$ < $\sim$18 kK). This may imply the termination of the MS phase at these effective temperatures within the luminosity range of 3.5 $\lesssim$ log($L/L_\odot$) $\lesssim$ 6.

    \item We detected only one SB2 system and a few lpv/sb1 targets among the most luminous BSGs with log($L/L_\odot$) > 5.

    \item We modeled the ability of the BLOeM observational campaign to detect different configurations of binaries within the available data for now and we estimated the corrected spectroscopic binary fraction to be 40 $\pm$ 4\%.   

    \item The observed spectroscopic binary fraction of BSGs in our sample is the same as in the Galaxy and the LMC; thus, within the present work analysis and data limitations, we did not detect any metallicity dependency regarding the number of binaries in the BSG domain.

    \item We made preliminary estimates of the orbital periods for 30 SB1 and SB2 systems within the 2 < $P_\mathrm{orb}$ < 100 d range. The Roche-lobe filling factor, together with eclipsing binary nature for some of these systems, could suggest that they are presently interacting binaries.

\end{itemize}

To summarize our results, the identification of various binary systems in our sample with a wide range of orbital periods and their non-uniform distribution across the HRD suggest that the BSG domain consists of objects that have different histories of binary interaction. Thus, phenomenologically, the BSG domain consists of various objects with different evolutionary paths, however, at the same time, in a morphological sense (observational classification), we define them as one class of objects. 

The continuation of the BLOeM campaign, including new observations and detailed determination of fundamental physical parameters based on the available spectra, will allow us to describe  given samples of BSG more extensively. This is especially promising with regard to the nature of the apparently single stars and detections of  long-period binaries. 

\begin{acknowledgements}
We thank the anonymous referee for helpful comments that have improved the manuscript. This project has received funding from the European Research Council (ERC) under the European Union's Horizon 2020 research and innovation programme (grant agreement 101164755/METAL) and was supported by the Israel Science Foundation (ISF) under grant number 2434/24. NB acknowledges support from the Belgian federal government grant for Ukrainian postdoctoral researchers (contract UF/2022/10). TS acknowledges support by the Israel Science Foundation (ISF) under grant number 0603225041. DP acknowledges financial support from the Deutsches Zentrum f\"ur Luft und Raumfahrt (DLR) grant FKZ 50OR2005 and the FWO junior postdoctoral fellowship No. 1256225N. DMB gratefully acknowledges UK Research and Innovation (UKRI) in the form of a Frontier Research grant under the UK government’s ERC Horizon Europe funding guarantee (SYMPHONY; PI Bowman; grant number: EP/Y031059/1), and a Royal Society University Research Fellowship (PI Bowman; grant number: URF{\textbackslash}R1{\textbackslash}231631). KS is funded by the National Science Center (NCN), Poland, under grant number OPUS 2021/41/B/ST9/00757. IM acknowledges support from the Australian Research Council (ARC) Centre of Excellence for Gravitational Wave Discovery (OzGav), through project number CE230100016. JIV acknowledges support from the European Research Council through ERC Advanced Grant No. 101054731. SS-D, and GH acknowledge support from the Spanish Ministry of Science and Innovation and Universities (MICIU) through the Spanish State Research Agency (AEI) through grants PID2021-122397NB-C21, and the Severo Ochoa Program 2020-2023 (CEX2019-000920-S).
\end{acknowledgements}


\bibliographystyle{aa}
\bibliography{ref}

\begin{appendix}
\section{Uncertainties in RV measurements and example of the line profile variability classification}

\begin{figure}[!ht]
\centering
\includegraphics[width=0.49\textwidth]{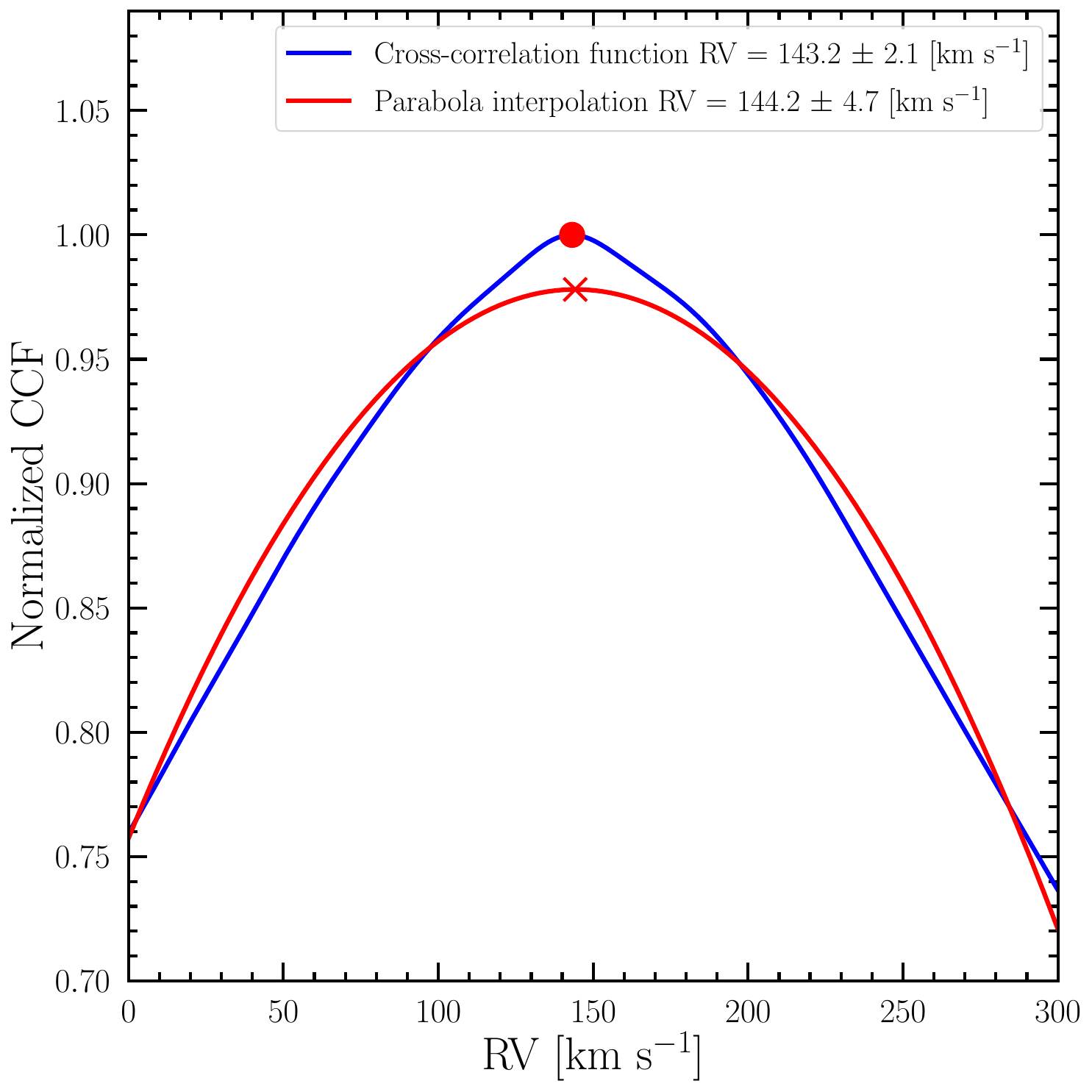}
\caption{Example of the computed cross-correlation function for one of the spectra with a relatively low S/N ratio (S/N $\sim$40, BLOeM 8-115$\_$01). The final radial velocity measurement (RV = 144.2 \kms) and its associated uncertainty  ($\sigma$RV = 4.7 \kms) are derived from the maximum and the sharpness of the parabola interpolation of CCF.}
\label{ccf_plot}
\end{figure}

Figure~\ref{ccf_plot} presents the example of CCF and its parabola fitting for one of the low S/N spectra in our sample. The CCF was computed by cross-correlating the full spectrum range with the stacked spectrum of all available epochs for a given star (usually 9 epochs). The stacked spectrum has been previously shifted to the zero-point of RV (i.e. the spectrum was fitted to the laboratory wavelengths of available absorption lines).  The CCF normalized to its maximum value, and it is well seen the non-symmetric top of the CCF which is caused by a noise in the spectrum. It is well seen that RV uncertainty based on parabola fitting is approximately twice larger than the uncertainty based on the original CCF, which is caused by the sharpness of the parabola fitting function. 

\begin{figure}[!ht]
\centering
\includegraphics[width=0.49\textwidth]{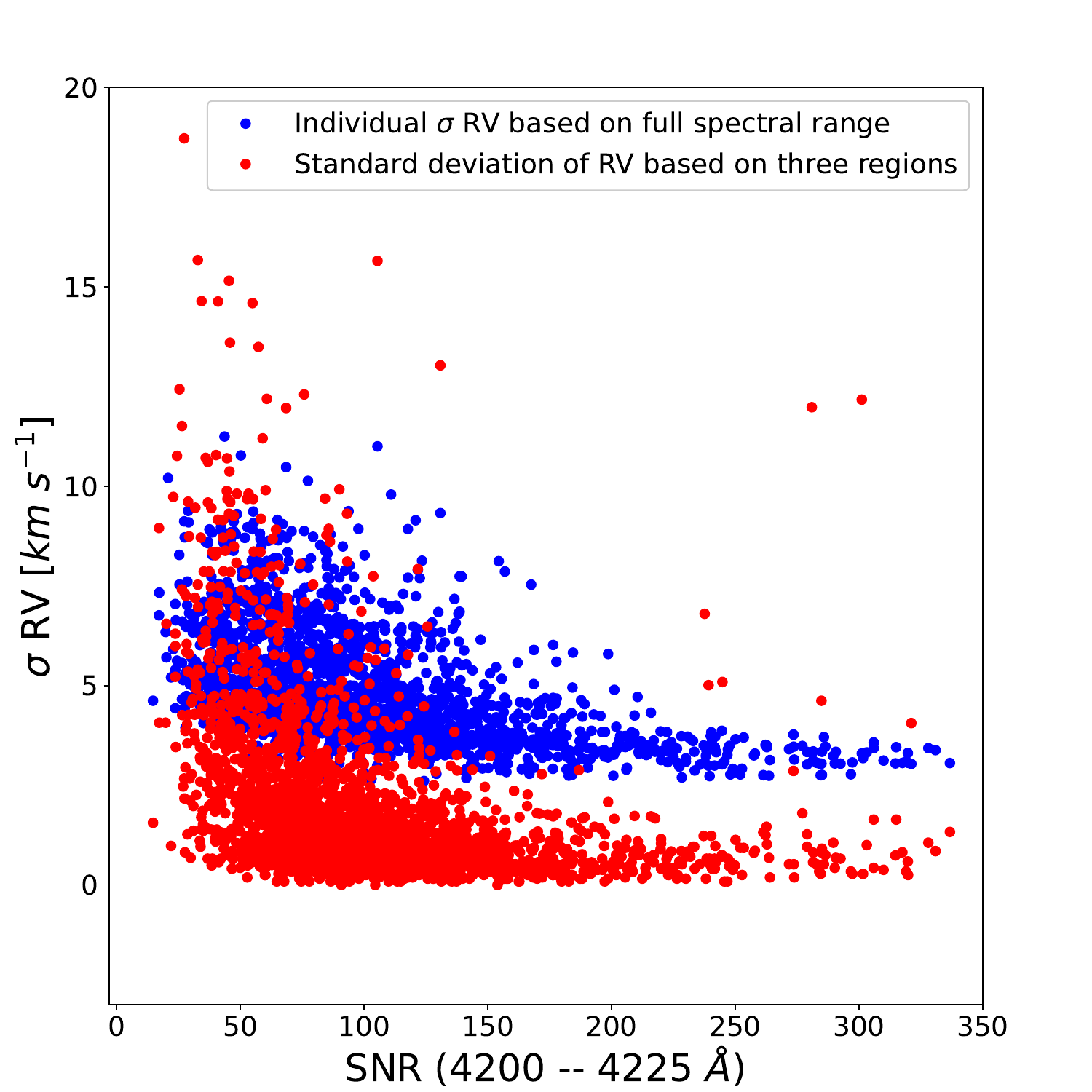}
\caption{Individual uncertainties of RV based on the cross-correlation of full spectral range (blue dots) and standard deviation of the mean RV  based on individual RV derived from three different wavelength regions (red dots) as a function of S/N for each analyzed spectrum.}
\label{errors_estimate}
\end{figure}

Figure~\ref{errors_estimate} illustrates estimates of the RV error ($\sigma$) as a function of S/N for each available BSG spectrum. We computed the S/N in a specific wavelength region (4200 -- 4225 \AA) which is free of any absorption lines. Individual RV errors are systematically higher than the standard deviation of three independent RV estimates (red dots), actually, it indicates that the parabola fitting of CCF is slightly overestimating the RV uncertainty as we discussed in the previous paragraph. The SB2 systems cause the most of high $\sigma$ outliers at S/N > 100.

\begin{figure*}[!ht]
\centering
\includegraphics[width=0.32\textwidth]{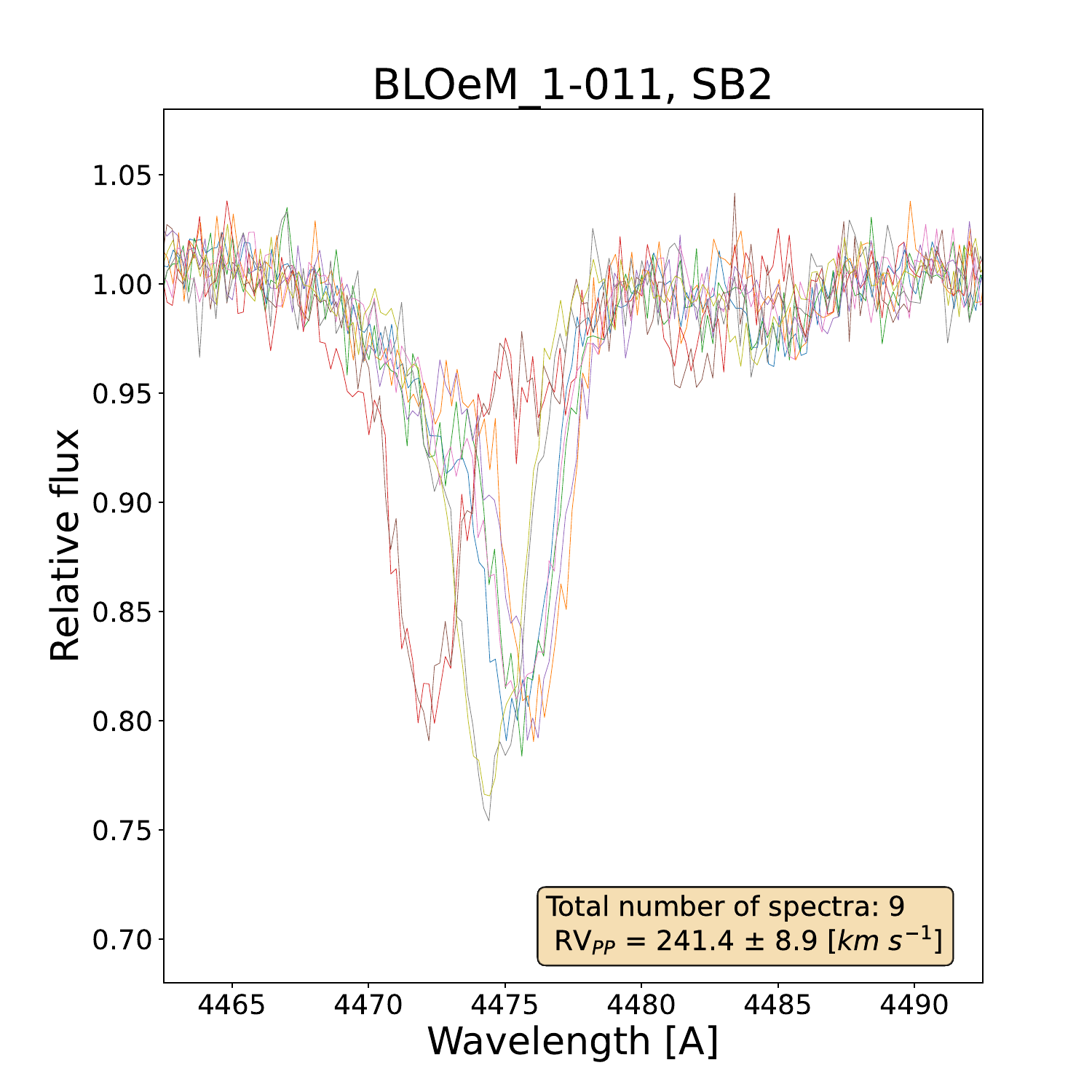}
\includegraphics[width=0.40\textwidth]{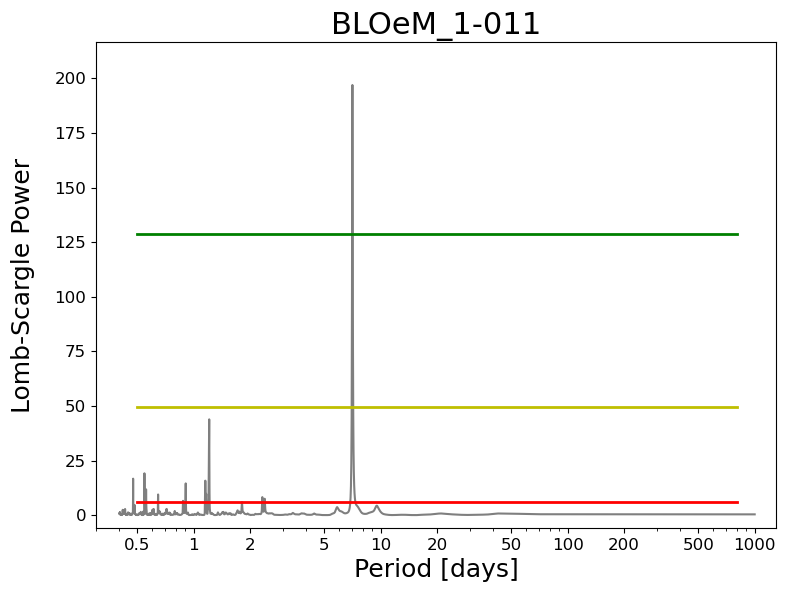}
\includegraphics[width=0.32\textwidth]{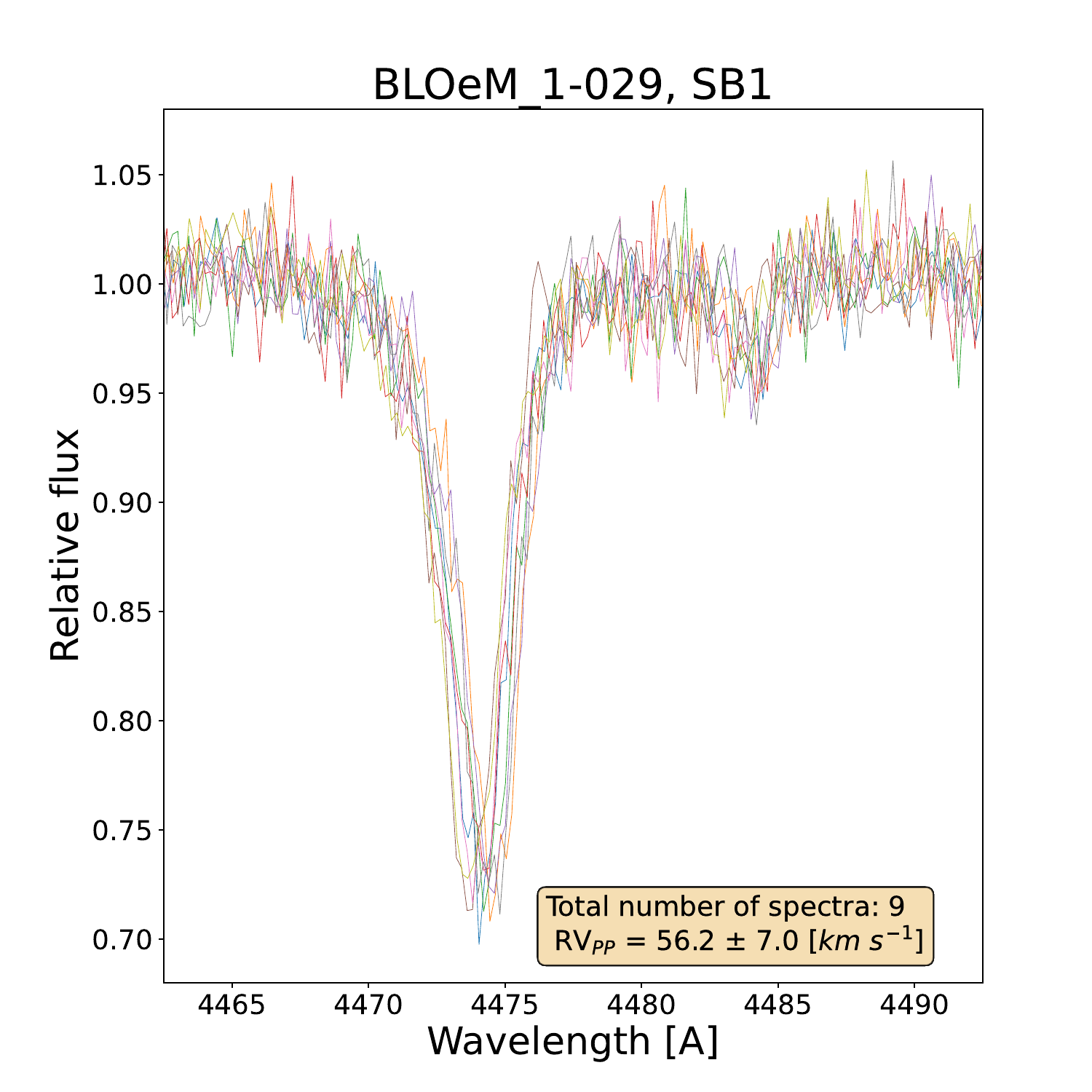}
\includegraphics[width=0.40\textwidth]{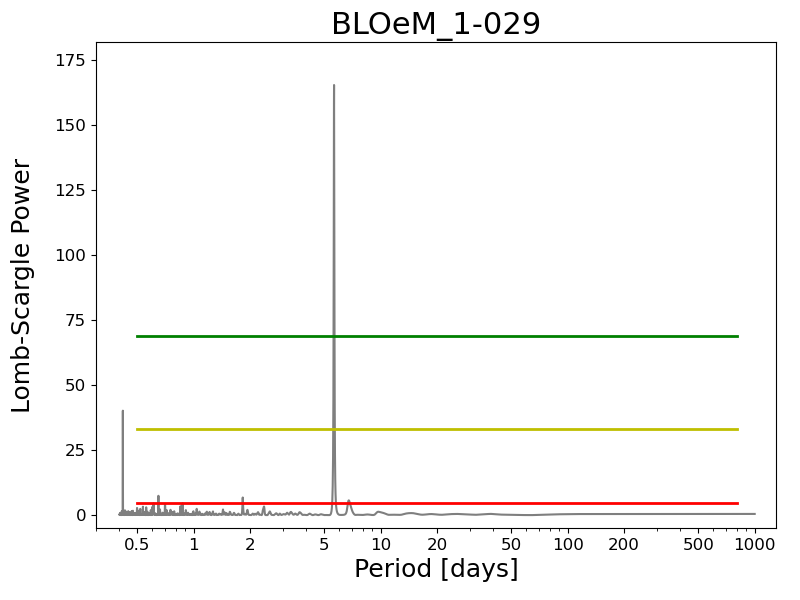}
\includegraphics[width=0.32\textwidth]{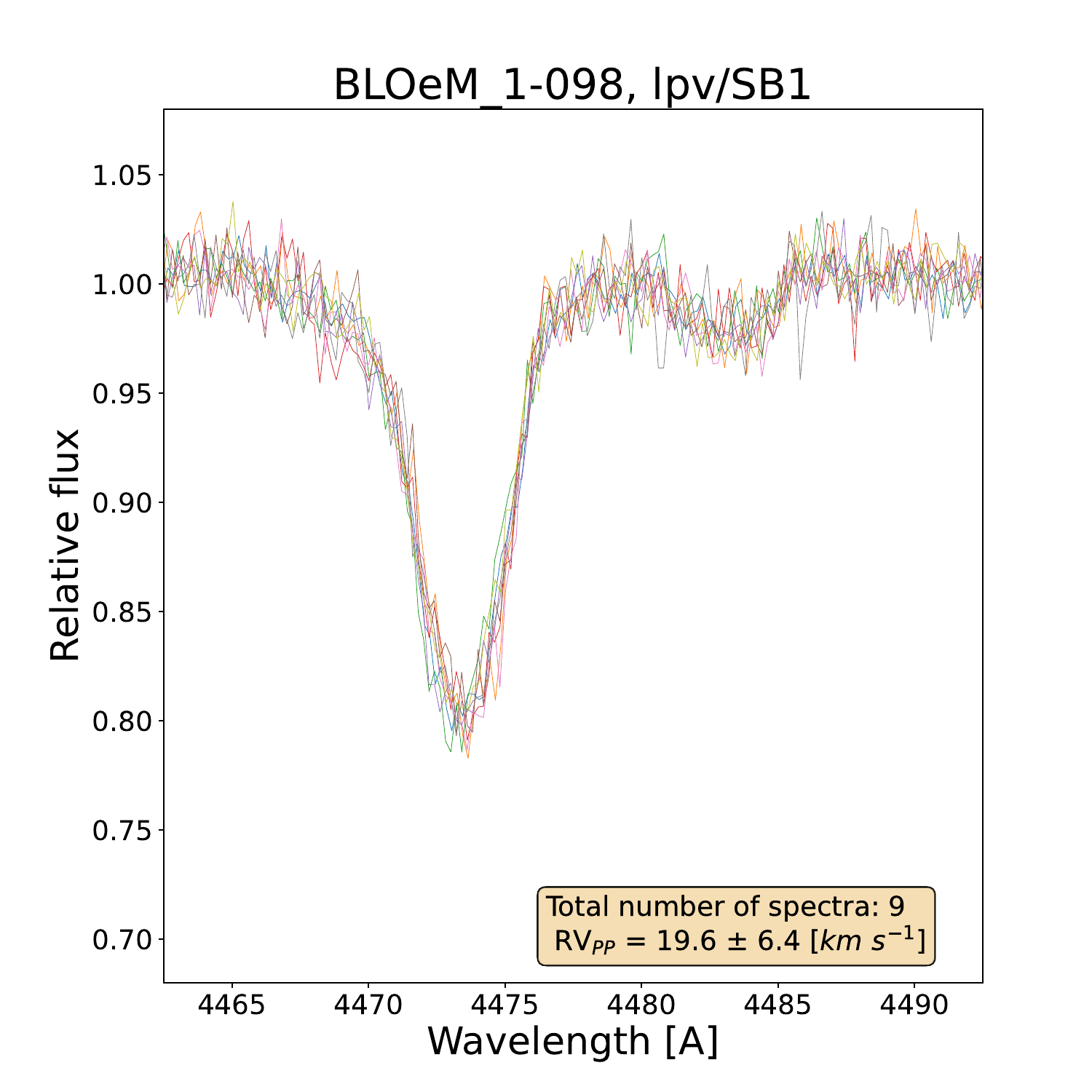}
\includegraphics[width=0.40\textwidth]{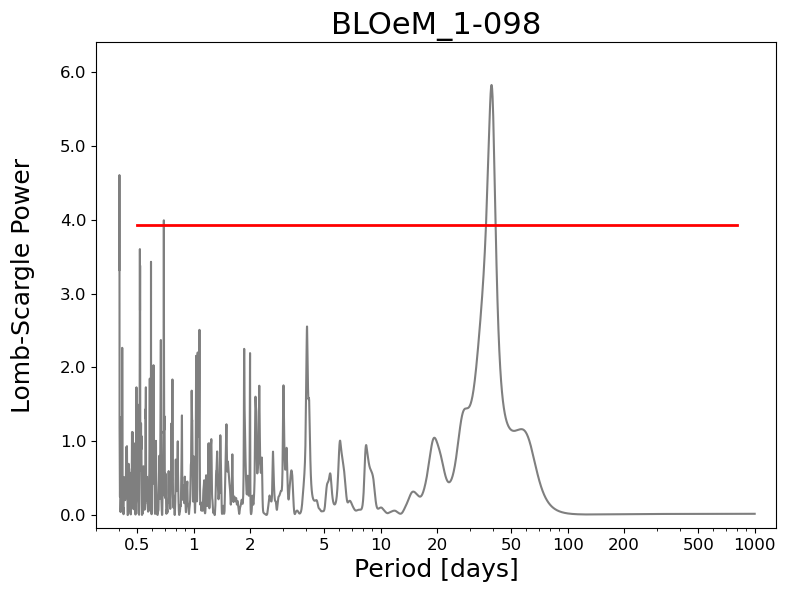}
\includegraphics[width=0.32\textwidth]{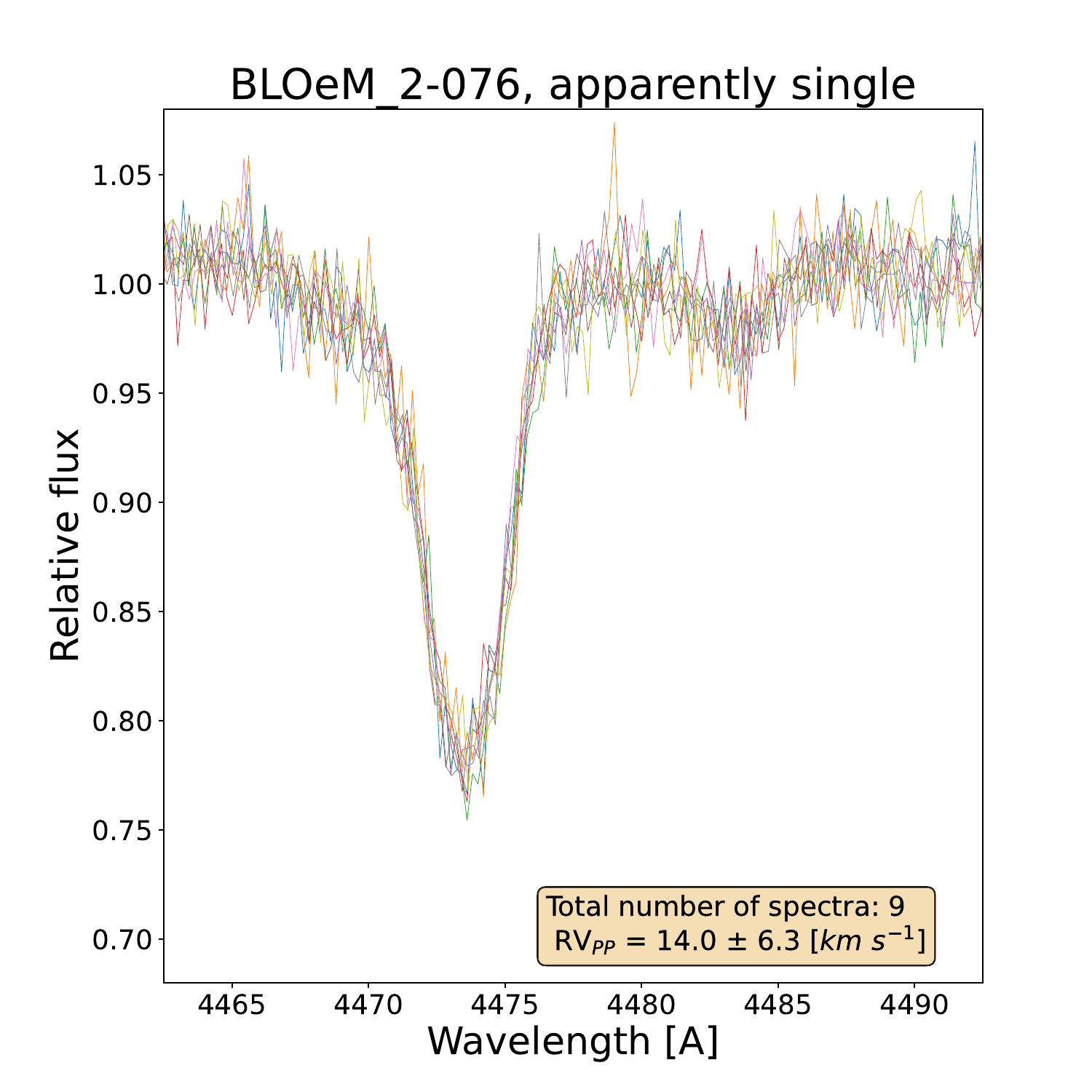}
\includegraphics[width=0.40\textwidth]{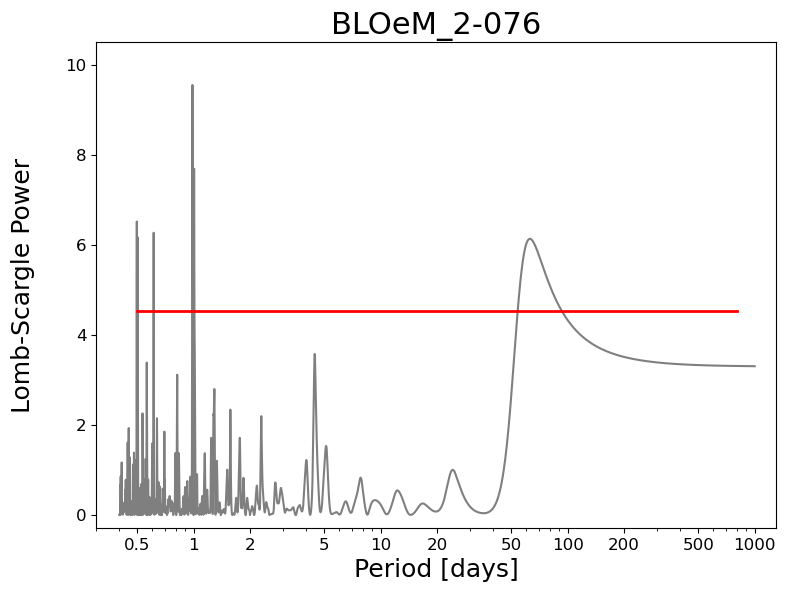}
\caption{Example of line profile variability for targets which we classified as SB2, SB1, lpv/SB1, and apparently single (left panels) and corresponded LS periodograms (right panels). Dashed horizontal lines on the LS periodograms represent different false-alarm levels namely, 0.1\% -- green line, 1\% -- yellow line, 50\% -- red line.}
\label{lpv_examples}
\end{figure*}

In Fig.~\ref{lpv_examples} we show line profile variability of the He\,{\sc i}\,$\lambda 4471$ line for four BSGs that represent our classification in terms of spectroscopic binary status (i.e., SB2, SB1, lpv/SB1, apparently single). In the right panels we show corresponding Lomb-Scargle periodograms together with different false-alarm levels (namely, 0.1\% -- green line, 1\% -- yellow line, 50\% -- red horizontal line).

\section{Supplementary data}
\label{sec:info_all_targets}

We present basic information and our spectroscopic binary classification for all BLOeM BSGs in Table \ref{table:all_bsgs}, namely: BLOeM ID, GAIA DR3 ID, spectral type, number of available epochs, peak-to-peak RV variation, effective temperature, luminosity, estimated radii, possible periods of RV variations, evolutionary masses, spectroscopic binary status, and notes.
The estimates of effective temperature and luminosities are taken from \citet{bloem_i}, and preliminary estimates of evolutionary masses are taken from Bestenlehner et al. (in prep.). The "Notes" column explanation:"(ULLYSES)": these BSGs have been studied within the ULLYSES survey \citet{Bernini-Peron_2024};  "broad line profiles" -- possible fast equatorial rotation; "low S/N" -  low signal-to-noise ratio of the spectra i.e. below 50; "possible long period binary"  -- prominent peaks in long-period ranges (>200 days) on the Lomb-Scargle periodogram; "EB" -- eclipsing binary; those periods that are marked with "*" are from the OGLE-IV database \citep{Pawlak_2016}.


\onecolumn
\begin{centering}
\LTcapwidth=\linewidth
\fontsize{7.5pt}{7.5pt}\selectfont 
\setlength{\tabcolsep}{2pt}
\begin{longtable}{c c c c c c c c c c c c c} 
\caption{Basic information about working sample of BSGs: BLOeM ID, GAIA DR3 ID, spectral type, number of available epochs, peak-to-peak RV variation, effective temperature, luminosity, estimated radii, possible periods of RV variations, evolutionary masses, spectroscopic binary status, and notes.}
\label{table:all_bsgs} \\
\hline
BLOeM  & GAIA DR3 & SpT & \# & \RVpp & $\sigma$ \RVpp & $T_{\rm eff}$ & log($L/L_\odot$) & R & P & $M_{\rm evol}$ & Sp. binary & Notes \\
 ID &  ID &  & epochs & [\kms] &   [\kms] & [K] &  & [R$_{\odot}$] & [days] & [M$_{\odot}$] & status &   \\
\hline
1-003   &   4690519151032663296 &    B2.5 II:  & 9 &   13.6 & 8.0  & 17200 &  3.98  &  11.0  &   --   & 10.1 &   --   &  \\
1-004   &   4690501146531455744 &    B1 Ib  & 9 &   8.4 & 5.1 & 22350 &  5.10  &  23.6  &   --   & 16.4 &   --   &  \\
1-005   &   4690505510214025472 &    B1 II  & 9 &   5.2 & 6.6 & 22350 &  4.34  &  9.9  &   --   & 15.5 &   --   &  \\
1-007   &   4690506781524350336 &    B1.5: II:  & 3 &   6.4  & 6.4 & 20650 &  3.90  &  7.0  &  --  & 12.1 &   --   &  \\
1-009   &   4690520387983090432 &    B1 Ia  & 9 &   12.8 & 3.9  & 22350 &  5.41  &  33.8  &   --   & 24.2 &   --   & AV 264 (ULLYSES)  \\
1-011   &   4690506712804881792 &    B1.5 II  & 9 &   241.4 & 8.9 & 20650 &  4.63  &  16.1  &  {\bf 7.07*}  & 12.8 &    {\bf SB2 }    &   EB  \\
1-014   &   4690520074427341184 &    B1 II  & 9 &   4.8 & 6.1 & 22350 &  4.44  &  11.1  &   --   & 12.7 &   --   &  \\
1-017   &   4690507193841111296 &    B1 II:  & 9 &   5.2 & 5.7 & 22350 &  4.24  &  8.8  &   --   & 10.1 &   --   &  \\
1-019   &   4690501558844407936 &    B2.5 Ib  & 9 &   10.8 & 6.9  & 19850 &  4.62  &  17.3  &   --   & 12.5 &  --  & \\
1-021   &   4690519425910500352 &    B1 Ib  & 9 &   19.4  & 8.8 & 22550 &  4.89  &  18.2  &   --   & 12.9 &  --  & \\
1-026   &   4690519455996745216 &    B1 Ib  & 9 &   6.2 & 7.1 & 22550 &  4.52  &  11.9  &   --   & 13.9 &  --  & \\
1-029   &   4690519803867505024 &    B2 II  & 9 &   56.2 & 7.0 & 18950 &  4.29  &  12.9  &   {\bf 5.63}  & 11.4 &    {\bf SB1 }  &  \\
1-030   &   4690506266128220160 &    B1 II  & 9 &   13.4 & 7.9 & 22350 &  3.70  &  4.7  &   --   & 10.8 &  --  & \\
1-034   &   4690519803867499648 &    B1 II  & 9 &   13.6 & 6.7  & 22350 &  4.53  &  12.3  &  --  & 12.8 &   lpv/SB1   & Possible period $\sim$13.94 days \\
1-036   &   4690519593389669632 &    B1.5 Ib  & 9 &   4.6 & 5.3 & 20650 &  5.13  &  28.7  &   --   & 19.3 &  --  & \\
1-038   &   4690513653473762176 &    B2.5 Ib  & 9 &   3.4 & 5.8  & 17200 &  4.53  &  20.7  &   --   & 12.5 &  --  & \\
1-042   &   4690512485242543104 &    B2 Ib  & 9 &   8.4 & 6.5  & 18950 &  4.82  &  23.8  &   --   & 15.4 &  --  & \\
1-044   &   4690512485242552960 &    B1 II  & 9 &   24.4 & 7.7  & 22350 &  4.23  &  8.7  &  {\bf  1.91}  & 13.6 &    {\bf SB1/SB2 }    &  \\
1-045   &   4690501971160817792 &    B0.7 II  & 9 &   4.4 & 5.3 & 24300 &  4.85  &  15.0  &   --   & 19.7 &  --  & \\
1-049   &   4690503375593898624 &    B1.5 II:  & 9 &   151.4 & 10.7 & 20650 &  4.56  &  14.9  &   --   & 15.2 &    {\bf SB2 }    &  Possible periods: $\sim$0.65 or  $\sim$15 days\\
1-052   &   4690512588321738496 &    B1 II  & 9 &   30.4 & 11.6  & 22350 &  4.40  &  10.6  &  --  & 12.7 &   lpv/SB1   &  Broad line profiles, low S/N\\
1-053   &   4690502280398443776 &    B1 Ib  & 9 &   4.0 & 5.6  & 22350 &  4.89  &  18.6  &   --   & 16.5 &  --  & \\
1-054   &   4690512584031941504 &    B0 II  & 9 &   58.0 & 5.8  & 27200 &  4.82  &  11.6  &   {\bf 65.69}  & 16.9 &    {\bf SB1 }    &  \\
1-057   &   4690513344235943040 &    B1.5 II:  & 9 &   6.8 & 8.8  & 20650 &  4.39  &  12.2  &   --   & 12.7 &  --  & \\
1-060   &   4690508907539818112 &    B1.5 Ib  & 9 &   7.4 & 7.0 & 20650 &  4.78  &  19.2  &   --   & 13.9 &  --  & \\
1-061   &   4690512244724379264 &    B1.5 Ib:  & 9 &   10.6 & 6.4  & 20650 &  4.71  &  17.7  &   --   & 12.9 &  --  & \\
1-070   &   4690512347803574400 &    B1.5 II  & 9 &   2.6 & 6.4  & 20650 &  4.49  &  13.7  &   --   & 12.8 &  --  & \\
1-074   &   4690509117967865088 &    B1 II  & 8 &   27.2 & 7.8  & 22350 &  3.91  &  6.0  &   {\bf 8.11}    & 14.4 &    {\bf SB1 }    &  \\
1-084   &   4687505527110195072 &    B1 II  & 9 &   5.8 & 6.5  & 22350 &  4.35  &  10.0  &   --   & 13.0 &  --  & \\
1-087   &   4690509732152391680 &    B1 II:  & 9 &   3.6 & 6.0 & 22350 &  4.50  &  11.9  &   --   & 12.6 &  --  & \\
1-089   &   4690514340668263552 &    B1 II  & 6 &   4.8 & 6.0  & 22350 &  4.74  &  15.6  &   --   & 13.5 &  --  & \\
1-092   &   4690513997070898944 &    B2 II  & 9 &   93.2 & 6.8  & 18950 &  4.23  &  12.1  &   {\bf 50.72}  & 10.5 &    {\bf SB1 }    &  \\
1-094   &   4687505217872848000 &    B1.5 II:  & 9 &   5.6 & 6.5  & 20650 &  4.16  &  9.4  &   --   & 9.7 &  --  & \\
1-095   &   4690508632639753984 &    B1 Ia  & 9 &   5.8 & 4.8  & 22350 &  5.40  &  33.4  &   --   & 28.7 &  --  & \\
1-096   &   4690509495924424960 &    B1 II  & 9 &   4.8 & 7.2  & 22350 &  3.88  &  5.8  &   --   & 14.1 &  --  & \\
1-097   &   4690509942621639680 &    B2.5 Ib  & 9 &   5.8 & 5.0 & 17200 &  4.87  &  30.6  &  --  & 16.4 &  --  & \\
1-098   &   4690512966278545792 &    B1 Ib  & 9 &   19.6 & 6.4  & 20650 &  4.52  &  14.2  &   --   & 12.6 &   lpv/SB1   &  The periods could be $\sim$20 -- 40 days\\
1-099   &   4690508671311394816 &    B2: II:  & 9 &   46.4 & 12.3  & 18950 &  4.04  &  9.7  &   --  & 13.1 &   lpv/SB1   &  Broad line profiles, low S/N, possible period $\sim$74.46 days\\
1-100   &   4690508740030862208 &    B1 II  & 9 &   56.8 & 7.3  & 22350 &  4.58  &  13.0  &   {\bf 4.28}  & 17.3 &    {\bf SB1 }    &  \\
1-103   &   4690508774390594944 &    B1 II:  & 9 &   31.6 & 7.3  & 22350 &  4.61  &  13.5  &    {\bf 117.17}  & 13.5 &    {\bf SB1 }    &  \\
1-105   &   4687504874275456896 &    B2: II  & 9 &   39.8 & 11.9  & 21200 &  3.92  &  6.8  &   --   & 10.6 &   lpv/SB1   &  low S/N \\
1-108   &   4687504977354372864 &    B1 II:  & 8 &   5.6 & 5.8  & 22350 &  4.33  &  9.7  &   --   & 10.7 &  --  & \\
1-109   &   4690510286219053312 &    B1.5 Ib  & 9 &   3.2 & 5.2  & 20650 &  4.82  &  20.1  &   --   & 16.6 &  --  & \\
1-110   &   4690510286219047936 &    B1 Ib  & 9 &   3.8 &  5.8  & 23950 &  4.84  &  15.3  &   --   & 13.4 &  --  & \\
1-111   &   4687507584364583040 &    B3 Ia  & 9 &   9.2 & 4.3  & 15500 &  5.52  &  79.7  &   --   & 24.0 &  --  &AV 362 (ULLYSES) \\
1-115   &   4690510419368517760 &    B1 II:  & 9 &   150.8 & 7.5  & 22350 &  4.42  &  10.8  &   {\bf  9.91 }  & 15.7 &    {\bf SB1 }    &  \\
1-116   &   4687507554335030656 &    B1 II  & 8 &   3.2 & 9.3  & 22350 &  4.44  &  11.1  &   --   & 14.2 &  --  & \\
2-002   &   4688959356376632960 &    B2 II  & 8 &   52.0 & 9.5  & 18950 &  3.77  &  7.1  &   --   & 9.0 &   lpv/SB1   &  low S/N \\
2-003   &   4688981823353120896 &    B1 II  & 9 &   1.8 & 6.2  & 22350 &  4.55  &  12.6  &   --   & 12.8 &  --  & \\
2-004   &   4688965575504047744 &    B3 II:  & 9 &   5.4 & 7.4  & 15500 &  3.79  &  10.9  &   --   & 7.6 &  --  & \\
2-011   &   4688977837586212608 &    B1 II:  & 9 &   2.8 & 6.3  & 22350 &  4.03  &  6.9  &   --   & 12.9 &  --  & \\
2-013   &   4688982343048079488 &    B1 II  & 9 &   13.0 & 5.4 & 22350 &  4.29  &  9.3  &   --   & 12.6 &  --  & \\
2-014   &   4688964613419753984 &    B1 II  & 9 &   2.2 & 5.6  & 22350 &  4.26  &  9.0  &   --   & 13.0 &  --  & \\
2-015   &   4688978219879759360 &    B2.5 II:  & 9 &   15.0 & 9.5  & 17200 &  3.97  &  10.9  &   --   & 12.6 &  --  & \\
2-025   &   4688978048081047296 &    B3 II:  & 9 &   3.4 & 6.2  & 15500 &  3.88  &  12.1  &   --   & 10.3 &  --  & \\
2-026   &   4688965777314555264 &    B2 II  & 9 &   490.4 & 7.9  & 18950 &  4.41  &  14.9  &   {\bf  7.11*}  & 12.9 &    {\bf SB2 }    &   EB  \\
2-028   &   4688977979361594880 &    B1.5 Ib  & 9 &   41.4 & 6.7  & 20650 &  4.70  &  17.5  &   {\bf  289.24 }  & 16.3 &    {\bf SB1 }    &   Long period binary\\
2-030   &   4688978082440780672 &    B2 II  & 9 &   5.2 & 8.2  & 18950 &  4.12  &  10.6  &   --   & 12.7 &   lpv/SB1   &  Broad line profiles, low S/N\\
2-031   &   4688966189635061760 &    B1 II  & 9 &   157.2 & 5.4  & 22350 &  4.71  &  15.1  &       & 13.7 &    {\bf SB1 }    & Possible period $\sim$13.06 days\\
2-032   &   4688964853937446912 &    B1 II  & 9 &   3.2 & 6.3  & 22350 &  4.50  &  11.9  &   --   & 12.7 &  --  & \\
2-034   &   4688961383604048768 &    B1 II  & 9 &   3.6 & 5.7  & 22350 &  4.27  &  9.1  &   --   & 14.9 &  --  & \\
2-036   &   4688960627689838976 &    B2 II:  & 9 &   5.4 & 6.8  & 18950 &  3.96  &  8.9  &   --   & 9.1 &  --  & \\
2-037   &   4688961761561107456 &    B2.5 Ib  & 9 &   8.2 & 4.1  & 17200 &  4.92  &  32.5  &   --   & 19.8 &  --  & \\
2-038   &   4688967735823484800 &    B1.5 II  & 9 &   11.0 & 8.5  & 20650 &  4.51  &  14.0  &   --   & 15.6 &  --  & \\
2-039   &   4688961864640239488 &    B1 II-Ib  & 9 &   1.6 & 5.1  & 22350 &  4.99  &  20.8  &   --   & 17.5 &  --  & \\
2-040   &   4688984026675105280 &    B2 II  & 9 &   1.8 & 5.3  & 18950 &  4.45  &  15.6  &   --   & 13.4 &  --  & \\
2-041   &   4688960657713202304 &    B2 II  & 9 &   12.0 & 7.8  & 18950 &  3.65  &  6.2  &   --   & 7.4 &   lpv/SB1   &  low S/N \\
2-043   &   4688961795920796928 &    B1 Iab  & 9 &   8.4 & 4.5  & 22350 &  5.05  &  22.3  &   --   & 20.9 &  --  & \\
2-044   &   4688967465277409408 &    B2: II  & 9 &   14.2 & 12.5  & 18950 &  3.92  &  8.5  &   --   & 10.2 &   lpv/SB1   &  Broad line profiles, low S/N\\
2-046   &   4688967396557954432 &    B1 II  & 9 &   21.0 & 6.2  & 22350 &  4.55  &  12.6  &    {\bf 34.17 }  & 12.7 &    {\bf SB1 }    &  \\
2-047   &   4688979594268679296 &    B1 Ib  & 9 &   25.8 & 6.8  & 22350 &  4.59  &  13.1  &    {\bf 37.54 }  & 14.1 &    {\bf SB1 }    &  Broad line profiles\\
2-057   &   4688980831219164416 &    B0.7 II  & 9 &   1.6 & 5.5  & 22850 &  4.77  &  15.5  &   --   & 17.0 &  --  & \\
2-060   &   4688963307749203072 &    B1.5 Ib  & 9 &   2.8 & 5.1  & 20650 &  4.61  &  15.8  &   --   & 15.4 &  --  & \\
2-074   &   4688968049392837248 &    B3 II:  & 9 &   4.40 & 6.3 & 15500 &  4.01  &  14.0  &   --   & 9.7 &  --  & \\
2-076   &   4688966881161489280 &    B1 Ib  & 9 &   14.0 & 6.3  & 22350 &  4.40  &  10.6  &   --   & 12.8 &  --  &  Possible long period binary\\
2-077   &   4688963926224354816 &    B1 II  & 9 &   22.8 & 6.2  & 22350 &  4.43  &  10.9  &   --   & 14.2 &   lpv/SB1   &  \\
2-078   &   4688963651346132224 &    B0 II  & 9 &   1.6 & 5.5  & 27200 &  4.91  &  12.8  &   --   & 19.9 &  --  & \\
2-083   &   4688963582626684544 &    B2.5 Ia  & 9 &   6.4 & 4.5  & 17200 &  5.13  &  41.3  &   --   & 24.1 &  --  & \\
2-084   &   4688962448755796608 &    B2 II:  & 9 &   66.8 & 5.9  & 18950 &  4.22  &  11.9  &   --   & 10.4 &    {\bf SB1 }    &  Possible period $\sim$23.07 days \\
2-089   &   4688968118111968512 &    B2 II  & 9 &   170.0 & 6.8  & 18950 &  4.24  &  12.2  &   {\bf 6.35*}  & 10.0 &    {\bf SB1 }    &  EB \\
2-102   &   4688990555027643904 &    B0 II:  & 9 &   2.2 & 5.5  & 27200 &  4.98  &  13.9  &   --   & 19.2 &  --  & \\
2-106   &   4688990589387328384 &    B0.7 II  & 9 &   70.4 & 5.4  & 22850 &  4.97  &  19.5  &    {\bf  7.54 }  & 21.6 &    {\bf SB1 }    &  \\
2-110   &   4688986431859207296 &    B0 II  & 9 &   2.4 & 5.4  & 27200 &  4.81  &  11.4  &   --   & 17.5 &  --  & \\
2-112   &   4688987389594710272 &    B2 II:  & 9 &   59.2 & 10.3  & 18950 &  4.52  &  16.9  &   {\bf 2.95*}  & 13.7 &    {\bf SB1/SB2 }    &  EB \\
2-113   &   4685983420730442624 &    B2.5 Ia  & 9 &   7.6 & 3.9  & 17200 &  5.24  &  46.9  &   --   & 23.9 &  --  & \\
2-114   &   4688986569298097792 &    B1.5 II  & 9 &   6.8 & 5.5  & 20650 &  4.59  &  15.4  &   --   & 13.8 &  --  & \\
3-001   &   4685853682687097088 &    B2.5 Ib  & 9 &   4.0 & 4.3  & 17200 &  4.91  &  32.1  &   --   & 19.8 &  --  & \\
3-017   &   4685836571534016128 &    B0.7 II  & 9 &   4.4 & 4.9  & 22850 &  4.79  &  15.8  &   --   & 19.6 &  --  & \\
3-021   &   4685835536431612288 &    B1.5 II  & 9 &   3.2 & 5.5  & 20650 &  4.50  &  13.9  &   --   & 12.7 &  --  & \\
3-022   &   4685948412524815616 &    B2 II neb  & 9 &   16.0 & 7.7  & 18950 &  4.49  &  16.3  &   --   & 14.0 &  --  & \\
3-023   &   4685849078482253440 &    B3 II  & 9 &   3.0 & 6.9  & 15500 &  3.99  &  13.7  &   --   & 10.3 &  --  & \\
3-028   &   4685947038135411200 &    B0.5 II  & 9 &   3.2 & 4.5  & 24300 &  4.82  &  14.5  &   --   & 19.9 &  --  & \\
3-029   &   4685944117556288640 &    B1.5 Ib  & 9 &   8.0 & 6.0  & 20650 &  4.69  &  17.3  &   --   & 15.3 &  --  & \\
3-030   &   4685947141214597376 &    B1 II  & 9 &   6.4 & 6.5  & 22350 &  4.60  &  13.3  &   --   & 13.2 &  --  & \\
3-032   &   4685836708972886400 &    B1.5 II  & 9 &   4.4 & 5.1  & 20650 &  4.61  &  15.8  &   --   & 14.3 &  --  & \\
3-037   &   4685835884339219456 &    B3 Ia  & 9 &   17.6 & 4.6  & 15500 &  5.27  &  59.8  &   --   & 24.1 &    lpv/SB1  &   The period could be different $\sim$30 days\\
3-039   &   4685944151915964928 &    B3 Ib  & 9 &   12.0 & 5.4  & 15500 &  4.76  &  33.2  &  --  & 15.0 &   lpv/SB1   &   Possible long period binary\\
3-044   &   4685929750832616320 &    B1 Ib  & 9 &   2.8 & 4.6  & 22350 &  4.97  &  20.4  &   --   & 17.4 &  --  & \\
3-047   &   4685944250646464768 &    B1 II  & 9 &   9.2 & 6.1  & 22350 &  4.81  &  16.9  &   --   & 15.6 &  --  & \\
3-056   &   4685947652249342080 &    B1.5 Ib  & 9 &   166.0 & 5.9  & 20650 &  4.64  &  16.3  &  {\bf  7.14*}  & 13.8 &    {\bf SB1 }    &  EB \\
3-057   &   4685926181767968000 &    B1 II  & 9 &   32.2 & 6.2  & 22350 &  4.31  &  9.5  &  {\bf  4.69 }  & 11.4 &    {\bf SB1 }    &  \\
3-059   &   4685947897128639744 &    B1 II  & 9 &   37.2 & 7.9  & 22350 &  4.77  &  16.2  &   --   & 15.6 &    {\bf SB1 }    & possible period $\sim$1.06 day \\
3-061   &   4685943254211136000 &    B3 Ib  & 9 &   4.8 & 4.1  & 15500 &  4.86  &  37.3  &   --   & 16.7 &  --  & \\
3-064   &   4685926353526652160 &    B3 II  & 9 &   3.0 & 6.6  & 15500 &  4.06  &  14.8  &   --   & 8.3 &  --  & \\
3-065   &   4685943254211144832 &    B0.7 II  & 9 &   4.4 & 4.9  & 22850 &  4.70  &  14.3  &   --   & 17.5 &  --  & \\
3-069   &   4685931468820884864 &    B1.5 II:  & 9 &   3.4 & 5.5  & 20650 &  4.44  &  13.0  &   --   & 12.4 &  --  & \\
3-070   &   4685926662804412800 &    B1 II  & 9 &   25.2 & 5.4  & 22350 &  4.38  &  10.3  &   {\bf  24.75 }  & 11.2 &    {\bf SB1 }    &  \\
3-073   &   4685943219851406208 &    B3 Ib  & 9 &   8.2 & 6.4  & 15500 &  4.78  &  34.0  &   {\bf 293.74*}  & 14.6 &  --  & EB (long-period system?) \\
3-074   &   4685926800243109760 &    B1 II  & 9 &   6.2 & 6.8  & 23950 &  4.57  &  11.2  &  --  & 13.0 &  --  & Possible long period binary\\
3-082   &   4685931335729612288 &    B1 II  & 9 &   137.0 & 7.6  & 22350 &  4.68  &  14.6  &   {\bf  7.38 }  & 13.0 &    {\bf SB2 }    &  \\
3-084   &   4685932744479939584 &    B0 Ib:  & 9 &   13.0 & 3.5  & 27200 &  4.98  &  13.9  &   --   & 16.8 &    {\bf SB1/SB2 }    &  Broad line profiles\\
3-090   &   4685932843197054208 &    B0.2 Ia  & 9 &   9.0 & 4.8  & 25750 &  5.21  &  20.2  &   --   & 24.3 &  --  & \\
3-093   &   4685925833826106368 &    B1 II:  & 9 &   11.8 & 6.6  & 22340 &  3.88  &  5.8  &    -- & 12.6 &   lpv/SB1   &  low S/N, possible period $\sim$12.49 days  \\
3-096   &   4685928410806791936 &    B2.5 II  & 9 &   3.6 & 5.1  & 17200 &  4.51  &  20.2  &   --   & 13.8 &  --  & \\
3-097   &   4685928415151555712 &    B2.5 II  & 9 &   10.6 & 6.7  & 17200 &  4.37  &  17.2  &   --   & 12.5 &  --  & \\
3-100   &   4685928449511258112 &    B3 II  & 9 &   3.2 & 6.9 & 15500 &  3.97  &  13.4  &   --   & 9.8 &  --  & \\
3-103   &   4685932126004485376 &    B1 II-Ib  & 9 &   6.4 & 6.9  & 22350 &  4.46  &  11.3  &   --   & 14.2 &  --  & \\
3-109   &   4685931816766886656 &    B0.2 Ib  & 9 &   3.0 & 4.9  & 25750 &  5.08  &  17.4  &   --   & 24.3 &  --  & \\
3-111   &   4685931812407476864 &    B1 Ib  & 9 &   2.6 & 4.8  & 22350 &  4.94  &  19.7  &   --   & 16.6 &  --  & \\
3-113   &   4685927453078935680 &    B2 II:  & 9 &   3.4 & 6.0  & 18950 &  4.44  &  15.4  &   --   & 11.0 &  --  & \\
3-115   &   4685928861827989120 &    B1 II  & 9 &   14.8 & 6.2  & 22350 &  4.66  &  14.2  &   5.07*  & 13.7 &   lpv/SB1   &  EB \\
3-116   &   4685928071554123264 &    B3 II  & 9 &   1.6 & 6.2 & 15500 &  4.04  &  14.5  &   --   & 9.9 &  --  & \\
4-003   &   4689019073615200768 &    B1 II  & 9 &   3.8 & 6.2  & 22350 &  4.25  &  8.9  &   --   & 11.9 &  --  & \\
4-011   &   4689014808648096896 &    B1 II  & 5 &   13.6 & 5.7  & 22350 &  4.33  &  9.7  &   --   & 12.9 &   lpv/SB1   &  Possible long period binary\\
4-015   &   4689019898248677888 &    B1 II-Ib  & 9 &   3.6 & 5.9  & 22350 &  4.55  &  12.6  &   --   & 12.9 &  --  & \\
4-017   &   4689002787095701248 &    B1.5 Ib  & 9 &   17.2 & 9.0  & 20650 &  4.59  &  15.4  &   {\bf  2.05 }  & 12.7 &   lpv/SB1   &  \\
4-019   &   4689002718376234112 &    B1 II  & 9 &   15.6 & 8.0  & 22350 &  4.44  &  11.1  &   --   & 12.7 &   lpv/SB1   &  Possible long period binary\\
4-020   &   4689003027613801216 &    B1 Iab-Ib  & 9 &   10.0 & 4.8  & 22350 &  5.41  &  33.8  &   --   & 30.1 &  --  &AV 210 (ULLYSES) \\
4-023   &   4689015225324629120 &    B2 II:  & 9 &   2.2 & 6.0  & 18950 &  4.15  &  11.0  &   --   & 10.2 &  --  & \\
4-024   &   4689003126336982272 &    B1 II  & 9 &   3.8 & 6.6  & 22350 &  4.44  &  11.1  &   --   & 11.2 &  --  & \\
4-027   &   4689001721943908096 &    B2 II  & 9 &   10.8 & 9.3  & 18950 &  3.59  &  5.8  &   --   & 8.5 &   lpv/SB1   &  Broad line profiles, possible long period binary\\
4-028   &   4689001653225111680 &    B2 II  & 9 &   21.4 &  9.6 & 18950 &  4.06  &  9.9  &   --   & 10.5 &   lpv/SB1   &  Broad line profiles, low S/N\\
4-030   &   4689015259684292608 &    B1 Ia  & 9 &   11.2 & 4.7  & 22350 &  5.19  &  26.2  &   --   & 21.2 &   lpv/SB1   &  Possible long period binary\\
4-036   &   4689002409138503040 &    B1.5 II  & 9 &   46.2 & 6.4  & 20650 &  4.39  &  12.2  &  {\bf 36.71 }  & 13.3 &    {\bf SB1 }    &  \\
4-042   &   4688999041884354944 &    B2.5 Ib  & 9 &   3.8 & 5.4  & 17200 &  4.81  &  28.6  &   --   & 18.7 &  --  & \\
4-045   &   4690521109537837440 &    B1 Iab  & 9 &   4.4 & 6.9  & 22350 &  4.63  &  13.8  &   --   & 12.9 &  --  & \\
4-046   &   4688998835725922176 &    B2 II:  & 9 &   13.2 & 8.0  & 18950 &  4.69  &  20.5  &   --   & 12.7 &   lpv/SB1   &  \\
4-047   &   4690518085880908544 &    B2 II  & 9 &   5.8 & 6.4 & 18950 &  4.10  &  10.4  &   --   & 10.1 &  --  & \\
4-048   &   4689003371211130624 &    B1 II  & 9 &   3.2 & 6.1  & 22350 &  4.39  &  10.4  &   --   & 12.1 &  --  & \\
4-050   &   4690521487494893952 &    B1 II:  & 8 &   7.8 & 6.1   & 22350 &  3.77  &  5.1  &   --   & 11.8 &  --  & \\
4-051   &   4688999140607234944 &    B1 II  & 9 &   3.4 & 6.6  & 22350 &  4.55  &  12.6  &   --   & 12.5 &  --  & \\
4-061   &   4689002271699550720 &    B2 II  & 9 &   28.2 & 11.1 & 18950 &  3.92  &  8.5  &   --   & 9.0 &   lpv/SB1   &  Broad line profiles, low S/N\\
4-065   &   4690517089448534272 &    B2 II:  & 9 &   52.4 & 8.5  & 18950 &  4.34  &  13.7  &  --  & 12.7 &    {\bf SB1 }    &  Possible long period binary\\
4-066   &   4690518223319786624 &    B2.5 Ib  & 9 &   3.4 & 5.0  & 17200 &  4.91  &  32.1  &   --   & 19.7 &  --  &AV 234 (ULLYSES) \\
4-067   &   4690518154600349184 &    B2 II:  & 4 &   79.4 & 10.8 & 18950 &  4.10  &  10.4  &   {\bf 3.75*}  & 9.2 &    {\bf SB2 }    &  EB \\
4-068   &   4690518154600325760 &    B2 II:  & 9 &   74.8 & 6.6  & 18950 &  4.23  &  12.1  &   {\bf 95.24}  & 10.9 &    {\bf SB1 }    &  EB (P = 1.39* days), possible triple system \\
4-069   &   4690517879722483328 &    B1 II neb  & 8 &   7.0 & 7.4  & 22350 &  3.70  &  4.7  &   --   & 10.0 &  --  & \\
4-070   &   4690516745851186048 &    B2 II neb  & 9 &   23.0 & 12.2  & 18950 &  3.55  &  5.5  &   --   & 7.7 &   lpv/SB1   &  Broad line profiles, possible long period binary\\
4-077   &   4690517055088782848 &    B3 Ib  & 9 &   206.2 & 4.3  & 15500 &  4.10  &  15.5  &   {\bf  49.17*}  & 10.4 &    {\bf SB1/SB2 }    &   EB \\
4-078   &   4690503826587160832 &    B1 Ia  & 9 &   31.6 & 4.8  & 22350 &  5.58  &  41.1  &   --   & 36.8 &  lpv/SB1    & AV 242  (ULLYSES) \\
4-083   &   4690500317576535040 &    B1 II:  & 9 &   34.2 & 6.2  & 22350 &  4.50  &  11.9  &   --  & 13.0 &    {\bf SB1 }    &  \\
4-085   &   4690505097897766144 &    B2.5 Ib  & 9 &   4.2 & 4.7  & 17200 &  4.88  &  31.0  &   --   & 16.6 &  --  & Possible long period binary\\
4-087   &   4690504341983098368 &    B1 II:  & 9 &   8.4 & 6.7  & 22350 &  4.22  &  8.6  &   --   & 14.1 &  --  & \\
4-090   &   4690503998385814272 &    B1 II  & 9 &   41.4 & 8.6  & 22350 &  4.40  &  10.6  &    {\bf 48.08 }  & 13.0 &    {\bf SB1 }    &  \\
4-096   &   4690504170184413952 &    B1.5 II:  & 9 &   7.2 & 9.8  & 20650 &  4.49  &  13.7  &   --   & 12.7 &  --  & Broad line profiles, low S/N\\
4-098   &   4690500802930242816 &    B2.5 II:  & 9 &   14.2 & 9.1  & 17200 &  4.52  &  20.5  &   {\bf 2.32}   & 10.8 &   lpv/SB1   &   Possible long period binary\\
4-101   &   4690499978296571392 &    B1.5 II:  & 9 &   3.4 & 5.3  & 20650 &  4.30  &  11.0  &  --   & 12.6 &  --  & \\
4-104   &   4690518841795059840 &    B1.5 II:  & 8 &   4.6 & 5.8  & 20650 &  4.37  &  12.0  &   --   & 10.7 &  --  & \\
4-105   &   4690501180887186944 &    B2: Ib:  & 9 &   13.0 & 6.0  & 18950 &  4.92  &  26.7  &   --   & 15.5 &  --  & \\
4-107   &   4690506747164583424 &    B1 II  & 9 &   4.8 & 6.9  & 22350 &  4.18  &  8.2  &   --   & 11.0 &  --  & \\
4-111   &   4690518979233951104 &    B3: II  & 9 &   95.0 & 12.9  & 15500 &  3.66  &  9.4  &   {\bf  5.27*}  & 9.5 &    {\bf SB1/SB2 }    &   EB \\
5-006   &   4687487763123080704 &    B2 II:  & 9 &   21.6 & 10.0  & 18950 &  3.97  &  9.0  &   {\bf  1.15 }  & 9.6 &    {\bf SB1 }    &  Broad line profiles, low S/N\\
5-007   &   4687499685954929664 &    B2.5 II  & 9 &   2.2 & 4.9  & 17200 &  4.45  &  18.9  &   --   & 12.8 &  --  & \\
5-008   &   4687501537095707264 &    B1 II:  & 9 &   4.0 & 6.8  & 22350 &  4.13  &  7.7  &   --   & 12.6 &  --  & \\
5-009   &   4687500063912170112 &    B2.5 II  & 9 &   5.2 & 6.2  & 17200 &  4.04  &  11.8  &   --   & 11.4 &  --  & Possible long period binary\\
5-010   &   4687501468343335808 &    B3 II  & 9 &   3.2 & 7.0  & 15500 &  3.92  &  12.6  &   --   & 9.6 &  --  & Possible long period binary\\
5-016   &   4687486556213576960 &    B1.5 II  & 9 &   2.6 & 5.9  & 20650 &  4.36  &  11.8  &   --   & 11.2 &  --  & \\
5-018   &   4687505763370767104 &    B3 II  & 9 &   2.2 & 6.1  & 15500 &  4.20  &  17.4  &   --   & 10.9 &  --  & \\
5-021   &   4687487999358589824 &    B1 II  & 9 &   2.0 & 5.4  & 22350 &  4.51  &  12.0  &   --   & 12.6 &  --  & \\
5-022   &   4687500579308038016 &    B3 II  & 9 &   5.6 & 6.0  & 15500 &  4.00  &  13.9  &   --   & 11.7 &  --  & \\
5-026   &   4687500304430157312 &    B1 II  & 9 &   2.2 & 5.5  & 22350 &  4.44  &  11.1  &   --   & 12.6 &  --  & \\
5-032   &   4687500270070426496 &    B2.5 II  & 9 &   2.0 & 6.3  & 17200 &  3.88  &  9.8  &   --   & 10.0 &  --  & \\
5-040   &   4687489927786395776 &    B1 II  & 9 &   71.6 & 5.0  & 22350 &  4.58  &  13.0  &   {\bf 128.7*}  & 13.7 &    {\bf SB1 }    &  EB \\
5-041   &   4687482815320929152 &    B1 II  & 9 &   3.6 & 5.2  & 22350 &  4.64  &  13.9  &   --   & 13.2 &  --  & \\
5-043   &   4687482712241716608 &    B1.5 II  & 9 &   2.6 & 5.4  & 20650 &  4.57  &  15.0  &   --   & 12.7 &  --  & \\
5-053   &   4687485319263467520 &    B3 II:  & 8 &   4.4 & 9.0  & 15500 &  3.56  &  8.4  &   --   & 8.3 &  --  & \\
5-056   &   4687501988057218688 &    B2 II  & 9 &   11.8 & 7.6  & 18950 &  4.18  &  11.4  &   --   & 10.3 &   lpv/SB1   &  low S/N \\
5-060   &   4687502572172727808 &    B1.5 II  & 9 &   2.2 & 5.6  & 20650 &  4.46  &  13.3  &   --   & 12.6 &  --  & \\
5-064   &   4687485564099788544 &    B1.5 II  & 9 &   1.0 & 5.1  & 20650 &  4.81  &  19.8  &   --   & 14.2 &  --  & \\
5-072   &   4687484361509061504 &    B1 II  & 9 &   2.6 & 5.3  & 22350 &  4.63  &  13.8  &   --   & 12.8 &  --  & \\
5-073   &   4687490305743468416 &    B2 II:  & 9 &   12.0 & 6.7  & 18950 &  4.06  &  9.9  &   --   & 12.6 &  --  & \\
5-075   &   4687502434733757440 &    B2 II:  & 9 &   42.6 & 13.2  & 18950 &  4.12  &  10.6  &   {\bf 2.95*}  & 12.6 &    {\bf SB2 }    &  EB \\
5-077   &   4687485491055825536 &    B2.5 Ia  & 9 &   7.0 & 4.3  & 17200 &  5.20  &  44.8  &   --   & 23.9 &  --  & \\
5-083   &   4687489584188971392 &    B1 II  & 8 &   6.4 & 6.3  & 22350 &  3.94  &  6.2  &   --   & 12.7 &  --  & \\
5-084   &   4687484773825841408 &    B2 II:  & 9 &   10.4 & 9.0  & 18950 &  4.16  &  11.1  &   --   & 9.8 &  --  & \\
5-087   &   4687489240591624832 &    B1 II  & 9 &   2.6 & 5.7  & 22350 &  4.30  &  9.4  &   --   & 11.1 &  --  & \\
5-093   &   4687437907140147712 &    B1 II  & 9 &   2.6 & 5.6  & 22350 &  4.53  &  12.3  &   --   & 12.5 &  --  & \\
5-103   &   4687509753348904832 &    B3 II  & 9 &   2.2 & 6.6  & 15500 &  4.04  &  14.5  &   --   & 10.5 &  --  & \\
5-104   &   4687512948804514176 &    B2.5 Ia  & 9 &   128.0 & 3.4  & 17200 &  4.77  &  27.3  &   {\bf  28.36 }  & 6.2 &    {\bf SB1 }    &  \\
5-105   &   4687509749039520768 &    B0.7 II  & 9 &   2.0 & 4.6  & 22850 &  5.06  &  21.6  &   --   & 19.8 &  --  & \\
5-106   &   4687508344599737216 &    B2 II  & 9 &   1.8  & 6.1  & 18950 &  4.29  &  12.9  &   --   & 12.7 &  --  & \\
5-107   &   4687512880085044608 &    B2.5 II  & 9 &   2.6 & 5.9  & 17200 &  4.47  &  19.3  &   --   & 13.5 &  --  & \\
5-113   &   4687508550758084864 &    B1 II  & 9 &   1.6 & 5.2  & 22350 &  4.83  &  17.3  &   --   & 14.3 &  --  & \\
5-114   &   4687508585117812736 &    B1.5 II  & 9 &   17.2 & 8.2  & 20650 &  4.66  &  16.7  &   --  & 12.8 &   lpv/SB1   & Possible period $\sim$1 day  \\
5-116   &   4687461649712663296 &    B2 II  & 8 &   15.2 & 8.8  & 18950 &  3.85  &  7.8  &    --  & 10.3 &   lpv/SB1     &  \\
6-017   &   4687165125163949952 &    B1 Ib  & 9 &   3.2 & 5.3  & 22350 &  5.12  &  24.2  &   --   & 21.2 &  --  & \\
6-069   &   4686413608956928640 &    B1 II  & 9 &   3.2 & 7.3  & 22350 &  3.95  &  6.3  &   --   & 12.8 &  --  & \\
6-070   &   4686414364871146624 &    B1: II  & 9 &   137.4 & 9.5  & 22350 &  4.62  &  13.6  &   {\bf  2.87*}  & 14.4 &    {\bf SB1 }    &  EB \\
6-072   &   4686413540237697408 &    B3 II  & 9 &   23.2 & 7.4  & 15500 &  4.22  &  17.9  &      --   & 9.9 &   lpv/SB1   & Possible period $\sim$9.55 days \\
6-076   &   4686414467950355072 &    B0.2 Ib  & 9 &   8.4 & 5.2  & 27200 &  5.52  &  25.9  &   --   & 29.2 &  --  & \\
6-078   &   4687165915437998848 &    B2 II  & 9 &   7.6 & 7.3  & 18950 &  3.66  &  6.3  &   --   & 8.6 &  --  & low S/N \\
6-080   &   4686413437158238208 &    B0.2 Ia  & 9 &   11.4 & 4.8  & 27200 &  5.90  &  40.1  &   --   & 46.0 &    lpv/SB1    &  AV 488 (ULLYSES) \\
6-099   &   4686414261791903360 &    B2 II  & 9 &   8.8 & 7.4  & 18950 &  3.55  &  5.5  &   --   & 8.2 &  --  & low S/N \\
6-100   &   4687168629857232000 &    B1 II  & 9 &   4.0 & 6.8  & 22350 &  4.36  &  10.1  &   --   & 11.7 &  --  & \\
6-102   &   4686416873131995776 &    B2 II  & 9 &   42.8 & 10.7  & 18950 &  3.85  &  7.8  &  --   & 8.8 &   lpv/SB1   &  Broad line profiles, possible long period. low S/N\\
6-111   &   4686416941851466880 &    B0.5 Ib  & 9 &   7.0 & 4.9  & 24300 &  5.38  &  27.6  &   --   & 31.2 &  --  & \\
7-002   &   4685991224768443392 &    B3 II:  & 9 &   4.0 & 6.6  & 15500 &  3.94  &  12.9  &   --   & 10.0 &  --  & \\
7-006   &   4688995468472128128 &    B1 II  & 9 &   3.8 & 5.7  & 22350 &  4.46  &  11.3  &   --   & 12.6 &  --  & Possible long period binary\\
7-007   &   4685987960593409024 &    B1 II neb  & 9 &   7.6 & 7.7  & 22350 &  4.50  &  11.9  &   --   & 12.9 &  --  & \\
7-008   &   4685992663601032576 &    B2 II:  & 8 &   13.0 & 10.2  & 18950 &  4.32  &  13.4  &   --   & 13.5 &  --  & Broad line profiles, low S/N\\
7-015   &   4685986792362398976 &    B1 II  & 9 &   2.6 & 6.2  & 22350 &  4.57  &  12.8  &   --   & 13.3 &  --  & Possible long period binary\\
7-019   &   4685986792362380928 &    B1.5: II  & 9 &   20.2 & 10.2  & 20650 &  4.17  &  9.5  &   --   & 12.5 &   lpv/SB1   &  Broad line profiles \\
7-022   &   4685988647787974784 &    B1 II:  & 9 &   5.4 & 6.5  & 22350 &  4.17  &  8.1  &  --   & 13.6 &  --  & \\
7-026   &   4685987479557047808 &    B1 II  & 9 &   5.4 & 6.7  & 22350 &  4.03  &  6.9  &  --  & 9.3 &  --  & \\
7-029   &   4685975453644517248 &    B2 II  & 9 &   8.6 & 9.5  & 18950 &  4.22  &  11.9  &   --   & 10.8 &  --  & Broad line profiles\\
7-035   &   4685975213126487424 &    B1 II:  & 9 &   154.0 & 6.7  & 22350 &  4.46  &  11.3  &   --   & 12.8 &    {\bf SB2 }    &  \\
7-041   &   4685994454583299840 &    B2 II:  & 9 &   16.6 & 9.9  & 18950 &  4.11  &  10.5  &   --   & 10.1 &  --  & \\
7-043   &   4685993423791257984 &    B1.5 Ib  & 9 &   3.4 & 4.9   & 20650 &  4.96  &  23.6  &   --   & 18.1 &  --  & \\
7-058   &   4685975934680882304 &    B1 II:  & 9 &   2.8 & 5.9   & 22350 &  4.45  &  11.2  &   --   & 13.7 &  --  & \\
7-059   &   4685972670505925888 &    B3 II  & 9 &   3.0 & 5.1  & 15500 &  4.09  &  15.4  &   --   & 9.9 &  --  & \\
7-061   &   4685976926752244480 &    B2: II:  & 9 &   21.4 & 12.3  & 18950 &  3.72  &  6.7  &   --   & 10.0 &   lpv/SB1   &  Broad line profiles, low S/N\\
7-064   &   4685972636146149504 &    B0 Ia  & 9 &   10.8 & 4.8  & 27200 &  5.89  &  39.6  &   --   & 45.9 &  --  &AV 235 (ULLYSES) \\
7-065   &   4685975762891732096 &    B1 II  & 8 &   57.0 & 9.9  & 22350 &  4.42  &  10.8  &    1.58*  & 12.8 &    {\bf SB1 }    &  EB \\
7-071   &   4685976209558662016 &    B0 II:  & 9 &   138.0 & 6.2  & 27200 &  5.11  &  16.2  &   {\bf  7.08 }  & 21.9 &    {\bf SB2 }    &  \\
7-073   &   4685989816018758272 &    B2 II:  & 9 &   10.8 & 9.7  & 18950 &  4.21  &  11.8  &   --   & 12.7 &  --  & Broad line profiles\\
7-083   &   4687490958580285440 &    B2 II:  & 9 &   6.6 & 7.9  & 18950 &  4.07  &  10.0  &   --   & 11.2 &  --  & \\
7-086   &   4687479070108329216 &    B2 II  & 9 &   8.8 & 9.0  & 18950 &  3.73  &  6.8  &  {\bf  18.11 }  & 8.9 &   lpv/SB1   &  \\
7-088   &   4685971948951359488 &    B2.5 II  & 9 &   3.0 & 5.9  & 17200 &  4.32  &  16.3  &   --   & 11.1 &  --  & \\
7-092   &   4687478863949902336 &    B2 II  & 9 &   7.4 & 8.4  & 18950 &  3.98  &  9.1  &   --   & 9.3 &  --  & Broad line profiles\\
7-096   &   4687490924220536576 &    B2 II:  & 9 &   15.6 & 9.1  & 18950 &  3.97  &  9.0  &   --   & 11.3 &   lpv/SB1   &  \\
7-098   &   4687490924220548736 &    B2 II:  & 9 &   20.0 & 9.0  & 18950 &  4.24  &  12.2  &   --   & 11.3 &   lpv/SB1   &  \\
7-099   &   4687477764438325376 &    B1 II  & 9 &   20.2 & 8.6  & 22350 &  4.55  &  12.6  &   --   & 14.4 &    {\bf SB2 }    &  Possible period ~1.56 days \\
7-100   &   4687479104468055552 &    B2 II  & 9 &   69.6 & 9.5  & 18950 &  3.68  &  6.4  &   --   & 9.2 &    {\bf SB1 }    &  \\
7-103   &   4687473641269604352 &    B1 II  & 9 &   5.0 & 5.5  & 22350 &  4.56  &  12.7  &   --   & 12.8 &  --  & \\
7-105   &   4687474397183820800 &    B1 II  & 9 &   7.4 & 7.1  & 22350 &  3.77  &  5.1  &   --   & 12.1 &  --  & \\
7-108   &   4687426087389568768 &    B2.5 Ia  & 9 &   15.0 & 5.8  & 17200 &  4.58  &  21.9  &  --  & 12.7 &    {\bf SB1 }    &   Possible long period\\
7-111   &   4687478138084449536 &    B0 Ia  & 9 &   17.4 & 5.1  & 27200 &  5.70  &  31.9  &   --   & 34.1 &    lpv/SB1     &  \\
7-112   &   4687478142395205888 &    B1 Ib  & 9 &   4.8 & 4.9  & 22350 &  4.86  &  17.9  &   --   & 16.5 &  --  & \\
7-113   &   4687475187457720448 &    B3 II  & 9 &   3.0 & 7.1  & 15500 &  4.13  &  16.1  &   --   & 12.2 &  --  & Possible long period binary\\
7-116   &   4687474942629350272 &    B3 II:  & 9 &   153.2 & 10.6  & 15500 &  3.65  &  9.3  &    2.53*  & 9.5 &    {\bf SB2 }    &   EB  \\
8-002   &   4689049894283437696 &    B2 II  & 9 &   160.0 & 7.8  & 18950 &  3.95  &  8.8  &   -- & 9.5 &    {\bf SB2 }    &  Possible period $\sim$1 day \\
8-006   &   4689037181200454656 &    B2.5 II-Ib  & 9 &   25.2 & 6.3  & 17200 &  4.64  &  23.5  &   --  & 13.9 &   lpv/SB1   &  Possible period $\sim$0.82 day \\
8-008   &   4689054945164735616 &    B1 Iab  & 9 &   11.2 & 4.6  & 22350 &  5.34  &  31.2  &   --   & 27.6 &  --  & \\
8-012   &   4689050890715677952 &    B2 II:  & 9 &   10.4 & 12.6  & 18950 &  3.88  &  8.1  &   --   & 10.9 &   lpv/SB1   &  Broad line profiles\\
8-013   &   4689038418150993920 &    B2 II  & 9 &   32.2 & 7.5  & 18950 &  3.94  &  8.7  &   {\bf  36.62 }  & 9.4 &    {\bf SB1 }    &  low S/N \\
8-018   &   4689075492291367936 &    B2 II:  & 9 &   11.8 & 9.1  & 18950 &  3.80  &  7.4  &   --   & 8.2 &  --  & \\
8-022   &   4689062298150064640 &    B0.7 II  & 9 &   6.8 & 5.5  & 24300 &  5.03  &  18.5  &   --   & 19.5 &  --  & \\
8-026   &   4689037456056165376 &    B2.5 II  & 9 &   3.8 & 6.1  & 17200 &  3.94  &  10.5  &   --   & 10.4 &  --  & \\
8-037   &   4688985985180031744 &    B2 II  & 9 &   5.0 & 7.6  & 18950 &  4.03  &  9.6  &   --   & 9.2 &  --  & \\
8-046   &   4689056766232136064 &    B1.5 II  & 9 &   3.2 & 5.7  & 20650 &  4.59  &  15.4  &   --   & 12.7 &  --  & \\
8-048   &   4689056456994593408 &    B3 II  & 9 &   3.2 & 6.3  & 15500 &  3.98  &  13.5  &   --   & 8.2 &  --  & \\
8-054   &   4689058071902092928 &    B2 II  & 7 &   2.6 & 6.6  & 18950 &  3.93  &  8.6  &   --   & 9.5 &  --  & \\
8-060   &   4689008177223333888 &    B2 II:  & 9 &   7.0 & 8.5  & 18950 &  4.28  &  12.8  &   --   & 12.5 &  --  & \\
8-063   &   4689008559534153856 &    B3 Ib  & 9 &   3.2 & 5.1  & 15500 &  4.60  &  27.6  &   --   & 15.7 &  --  & \\
8-066   &   4689058174981307392 &    B2 II:  & 9 &   11.8 & 6.6  & 18950 &  4.29  &  12.9  &   --   & 10.7 &  --  & Possible long period binary\\
8-078   &   4689010067007622656 &    B2 II:  & 9 &   16.4 & 8.0  & 18950 &  4.55  &  17.5  &    --  & 10.7 &   lpv/SB1   & Possible period $\sim$2.10 days \\
8-085   &   4689057041109984128 &    B1 II  & 9 &   2.6 & 5.9  & 22350 &  4.61  &  13.5  &   --   & 10.4 &  --  & \\
8-090   &   4689057487786300288 &    B1 II  & 9 &   121.2 & 8.0  & 22350 &  4.40  &  10.6  &   {\bf 10.92*}  & 12.8 &    {\bf SB1 }    &  EB \\
8-094   &   4689059824248453888 &    B3 Ib  & 9 &   3.4 & 5.1  & 15500 &  4.75  &  32.9  &   --   & 15.0 &  --  & \\
8-098   &   4689058656017416320 &    B3 II  & 9 &   3.8 & 6.4  & 15500 &  4.04  &  14.5  &   --   & 8.8 &  --  & \\
8-099   &   4689058724736877824 &    B2 II:  & 9 &   10.8 & 10.3  & 18950 &  3.99  &  9.2  &   --   & 9.0 &   lpv/SB1   &  Broad line profiles\\
8-115   &   4689058793456333824 &    B1 II  & 9 &   4.0 & 6.6  & 22350 &  4.11  &  7.6  &   --   & 12.7 &  --  & \\
\hline
\end{longtable}
{\bf Notes.} Those periods that are marked with "*" are from the OGLE-IV database \citep{Pawlak_2016}. 
\end{centering}

\twocolumn

\end{appendix}

\end{document}